\documentclass[aps,floatfix,prc,twocolumn,showpacs,superscriptaddress]{revtex4-1} 
\usepackage{longtable,dcolumn,epsfig,color}
\usepackage{amsmath}
\usepackage{url}
\usepackage{hyperref}
\usepackage{rotating}
\usepackage{afterpage}
\usepackage{xspace}
\usepackage{graphicx}

\begin{document}
\def\nuc#1#2{${}^{#1}$#2}
\def\mee{$\langle m_{\beta\beta} \rangle$}
\def\mnu{$m_{\nu}$}
\def\ml{$m_{lightest}$}
\def\gnu{$\langle g_{\nu,\chi}\rangle$}
\def\mmod{$\| \langle m_{\beta\beta} \rangle \|$}
\def\mb{$\langle m_{\beta} \rangle$}
\def\BBz{$\beta\beta(0\nu)$}
\def\BBm{$\beta\beta(0\nu,\chi)$}
\def\BBt{$\beta\beta(2\nu)$}
\def\nonubb{$0\nu\beta\beta$}
\def\twonubb{$2\nu\beta\beta$}
\def\BB{$\beta\beta$}
\def\Mz{$M_{0\nu}$}
\def\Mt{$M_{2\nu}$}
\def\MzG{$M^{GT}_{0\nu}$} 
\def\MzF{$M^{F}_{0\nu}$} 
\def\MtG{$M^{GT}_{2\nu}$} 
\def\MtF{$M^{F}_{2\nu}$} 
\def\Gz{$G_{0\nu}$}					
\def\Tz{$T^{0\nu}_{1/2}$}
\def\Tt{$T^{2\nu}_{1/2}$}
\def\Tc{$T^{0\nu\,\chi}_{1/2}$}
\def\Rz{$\Gamma_{0\nu}$} 
\def\Rt{$\Gamma_{2\nu}$} 
\def\ms{$\delta m_{\rm sol}^{2}$}
\def\ma{$\delta m_{\rm atm}^{2}$}
\def\mot{$\delta m_{12}^{2}$}
\def\mtt{$\delta m_{23}^{2}$}
\def\ts{$\theta_{\rm sol}$}
\def\ta{$\theta_{\rm atm}$}
\def\ttwo{$\theta_{12}$}
\def\tot{$\theta_{13}$}
\def\gpp{$g_{pp}$} 
\def\gA{$g_{A}$} 
\def\qval{$Q_{\beta\beta}$} 
\def\be{\begin{equation}}
\def\ee{\end{equation}}
\def\cpKkgy{cts/(keV kg yr)}
\def\cpKkgd{cts/(keV kg d)}
\def\cpRty{cts/(ROI ton yr)}
\def\onecpRty{1~cts/(ROI ton yr)}
\def\threecpRty{3~cts/(ROI ton yr)}
\def\ppc{P-PC} 
\def\nsc{N-SC} 
\def\cosixty{$^{60}Co$}
\def\thttt{$^{232}\mathrm{Th}$}
\def\utte{$^{238}\mathrm{U}$}
\def\mubqkg{$\mu\mathrm{Bq/kg}$}
\def\cusulfate{$\mathrm{CuSO}_4$}
\def\MJ{{\sc Majorana}} 
\def\DEM{{\sc Demonstrator}} 
\def\MJD{{\sc{Majorana Demonstrator}}}
\def\MJDEMbf{\bfseries{\scshape{Majorana Demonstrator}}}
\def\MJbf{\bfseries{\scshape{Majorana}}}
\def\MJDEMit{\itshape{\scshape{Majorana Demonstrator}}}
\newcommand{\Gerda}{\textsc{Gerda}}
\newcommand{\Geant}{\textsc{Geant}}
\newcommand{\MaGe}{\textsc{MaGe}}

\title{Signatures of muonic activation in the \MJD}

\newcommand{\ITEP}{National Research Center ``Kurchatov Institute'' Institute for Theoretical and Experimental Physics, Moscow, 117218 Russia}
\newcommand{\JINR}{Joint Institute for Nuclear Research, Dubna, 141980 Russia} 
\newcommand{\lbnl}{Nuclear Science Division, Lawrence Berkeley National Laboratory, Berkeley, CA 94720, USA}
\newcommand{\lbnle}{Engineering Division, Lawrence Berkeley National Laboratory, Berkeley, CA 94720, USA}
\newcommand{\lanl}{Los Alamos National Laboratory, Los Alamos, NM 87545, USA}
\newcommand{\queens}{Department of Physics, Engineering Physics and Astronomy, Queen's University, Kingston, ON K7L 3N6, Canada}
\newcommand{\uw}{Center for Experimental Nuclear Physics and Astrophysics, and Department of Physics, University of Washington, Seattle, WA 98195, USA}
\newcommand{\unc}{Department of Physics and Astronomy, University of North Carolina, Chapel Hill, NC 27514, USA}
\newcommand{\duke}{Department of Physics, Duke University, Durham, NC 27708, USA}
\newcommand{\ncsu}{Department of Physics, North Carolina State University, Raleigh, NC 27695, USA}	
\newcommand{\ornl}{Oak Ridge National Laboratory, Oak Ridge, TN 37830, USA}
\newcommand{\ou}{Research Center for Nuclear Physics, Osaka University, Ibaraki, Osaka 567-0047, Japan}
\newcommand{\pnnl}{Pacific Northwest National Laboratory, Richland, WA 99354, USA}
\newcommand{\ttu}{Tennessee Tech University, Cookeville, TN 38505, USA}
\newcommand{\sdsmt}{South Dakota Mines, Rapid City, SD 57701, USA}
\newcommand{\usc}{Department of Physics and Astronomy, University of South Carolina, Columbia, SC 29208, USA}
\newcommand{\usd}{Department of Physics, University of South Dakota, Vermillion, SD 57069, USA}  
\newcommand{\ut}{Department of Physics and Astronomy, University of Tennessee, Knoxville, TN 37916, USA}
\newcommand{\tunl}{Triangle Universities Nuclear Laboratory, Durham, NC 27708, USA}
\newcommand{\mpi}{Max-Planck-Institut f\"{u}r Physik, M\"{u}nchen, 80805, Germany}
\newcommand{\tum}{Physik Department and Excellence Cluster Universe, Technische Universit\"{a}t, M\"{u}nchen, 85748 Germany}
\newcommand{\williams}{Physics Department, Williams College, Williamstown, MA 01267, USA}
\newcommand{\ciemat}{Centro de Investigaciones Energ\'{e}ticas, Medioambientales y Tecnol\'{o}gicas, CIEMAT 28040, Madrid, Spain}
\newcommand{\iu}{Department of Physics, Indiana University, Bloomington, IN 47405, USA}
\newcommand{\iuceem}{IU Center for Exploration of Energy and Matter, Bloomington, IN 47408, USA}

\author{I.J.~Arnquist}\affiliation{\pnnl} 
\author{F.T.~Avignone~III}\affiliation{\usc}\affiliation{\ornl}
\author{A.S.~Barabash}\affiliation{\ITEP}
\author{C.J.~Barton}\affiliation{\usd}	
\author{F.E.~Bertrand}\affiliation{\ornl}
\author{E.~Blalock}\affiliation{\ncsu}\affiliation{\tunl} 
\author{B.~Bos}\affiliation{\unc}\affiliation{\tunl} 
\author{M.~Busch}\affiliation{\duke}\affiliation{\tunl}	
\author{M.~Buuck}\altaffiliation{Present address: SLAC National Accelerator Laboratory, Menlo Park, CA 94025, USA}\affiliation{\uw}
\author{T.S.~Caldwell}\affiliation{\unc}\affiliation{\tunl}	
\author{Y-D.~Chan}\affiliation{\lbnl}
\author{C.D.~Christofferson}\affiliation{\sdsmt} 
\author{P.-H.~Chu}\affiliation{\lanl} 
\author{M.L.~Clark}\affiliation{\unc}\affiliation{\tunl} 
\author{C.~Cuesta}\affiliation{\ciemat}	
\author{J.A.~Detwiler}\affiliation{\uw}	
\author{T.R.~Edwards}\affiliation{\lanl}\affiliation{\usd} 
\author{Yu.~Efremenko}\affiliation{\ut}\affiliation{\ornl}
\author{H.~Ejiri}\affiliation{\ou}
\author{S.R.~Elliott}\affiliation{\lanl}
\author{G.K.~Giovanetti}\affiliation{\williams}  
\author{M.P.~Green}\affiliation{\ncsu}\affiliation{\tunl}\affiliation{\ornl}   
\author{J.~Gruszko}\affiliation{\unc}\affiliation{\tunl} 
\author{I.S.~Guinn}\affiliation{\unc}\affiliation{\tunl} 
\author{V.E.~Guiseppe}\affiliation{\ornl}	
\author{C.R.~Haufe}\affiliation{\unc}\affiliation{\tunl}	
\author{R.~Henning}\affiliation{\unc}\affiliation{\tunl}
\author{D.~Hervas~Aguilar}\affiliation{\unc}\affiliation{\tunl} 
\author{E.W.~Hoppe}\affiliation{\pnnl}
\author{A.~Hostiuc}\affiliation{\uw} 
\author{M.F.~Kidd}\affiliation{\ttu}	
\author{I.~Kim}\affiliation{\lanl} 
\author{R.T.~Kouzes}\affiliation{\pnnl}
\author{T.E.~Lannen~V}\affiliation{\usc}
\author{A.M.~Lopez}\affiliation{\ut}	
\author{J.M. L\'opez-Casta\~no}\affiliation{\usd} 
\author{E.L.~Martin}\affiliation{\unc}\affiliation{\tunl}	
\author{R.D.~Martin}\affiliation{\queens}	
\author{R.~Massarczyk}\affiliation{\lanl}		
\author{S.J.~Meijer}\affiliation{\lanl}	
\author{S.~Mertens}\affiliation{\mpi}\affiliation{\tum}		
\author{T.K.~Oli}\affiliation{\usd}  
\author{G.~Othman}\affiliation{\unc}\affiliation{\tunl} 
\author{L.S.~Paudel}\affiliation{\usd} 
\author{W.~Pettus}\affiliation{\iu}\affiliation{\iuceem}	
\author{A.W.P.~Poon}\affiliation{\lbnl}
\author{D.C.~Radford}\affiliation{\ornl}
\author{A.L.~Reine}\affiliation{\unc}\affiliation{\tunl}	
\author{K.~Rielage}\affiliation{\lanl}
\author{N.W.~Ruof}\affiliation{\uw}	
\author{D.~Tedeschi}\affiliation{\usc}		
\author{R.L.~Varner}\affiliation{\ornl}  
\author{S.~Vasilyev}\affiliation{\JINR}	
\author{J.F.~Wilkerson}\affiliation{\unc}\affiliation{\tunl}\affiliation{\ornl}    
\author{C.~Wiseman}\affiliation{\uw}		
\author{W.~Xu}\affiliation{\usd} 
\author{C.-H.~Yu}\affiliation{\ornl}
\author{B.X.~Zhu}\altaffiliation{Present address: Jet Propulsion Laboratory, California Institute of Technology, Pasadena, CA 91109, USA}\affiliation{\lanl} 

\collaboration{{\sc{Majorana}} Collaboration}
\noaffiliation

\date{\today}

\begin{abstract}
Experiments searching for very rare processes such as neutrinoless double-beta decay require a detailed understanding of all sources of background. Signals from radioactive impurities present in construction and detector materials can be suppressed using a number of well-understood techniques. Background from in-situ cosmogenic interactions can be reduced by siting an experiment deep underground. However, the next generation of such experiments have unprecedented sensitivity goals of 10$^{28}$ years half-life with background rates of 10$^{-5}$\cpKkgy{} in the region of interest. To achieve these goals, the remaining cosmogenic background must be well understood. In the work presented here, \textsc{Majorana Demonstrator} data is used to search for decay signatures of meta-stable germanium isotopes. Contributions to the region of interest in energy and time are estimated using simulations, and compared to \DEM~data. Correlated time-delayed signals are used to identify decay signatures of isotopes produced in the  germanium detectors. A good agreement between expected and measured rate is found and different simulation frameworks are used to estimate the uncertainties of the predictions. The simulation campaign is then extended to characterize the background for the LEGEND experiment, a proposed tonne-scale effort searching for neutrinoless double-beta decay in $^{76}$Ge. 
\end{abstract}
\maketitle
\section{Introduction}
Interactions with cosmogenic particles are an important source of background for rare event searches such as dark matter~\cite{Kozlov2010,Schmidt2013, Davis2014, Mayet2016}, neutrino oscillations~\cite{Barker2012}, or neutrinoless double-beta decay (\nonubb)~\cite{Pandola2007, Bellini2010, Abe2010}. Therefore, these experiments are usually sited in laboratories deep underground to reduce the cosmic ray flux. However, even after a reduction by orders of magnitude, the remaining flux can be a problem for the next generation of underground experiments.
The first few hundred feet of rock overburden will completely absorb many types of cosmic rays, but high-energy muons can penetrate several thousand feet of rock. Muons with kinetic energies up into the TeV range can interact with rock or the experimental apparatus and create large numbers of secondary particles. These particle showers often have an electromagnetic component which includes photons, and can also have a hadronic component which includes protons or neutrons~\cite{PhysRevC.99.055810, Kudryavtsev2003, Lindote2009, Araujo2008, Malgin2015}. 

One such deep underground rare event search is the \MJD~(MJD)~\cite{Abgrall2014,Aalseth2018,Alvis_2019}. This \nonubb~experiment is located at the 4850-ft level of the Sanford Underground Research Facility (SURF) \cite{Heise2015} in Lead, South Dakota. At such depths, the muon flux is reduced by orders of magnitude relative to the surface. A recent measurement found $(5.31\pm 0.16)\times 10^{-9}~\mu$ cm$^{-2}$ s$^{-1}$~\cite{Abgrall2017} for the total muon flux. Because of the low-background nature of these experiments, complementary measurements and simulations are necessary in order to understand the contribution of the remaining cosmogenic flux~\cite{Wei:2017hkq,Wiesinger_2018,Du:2018mzh}.

\begin{figure}[t]
\centering
\includegraphics[width=0.9\columnwidth,keepaspectratio=true]{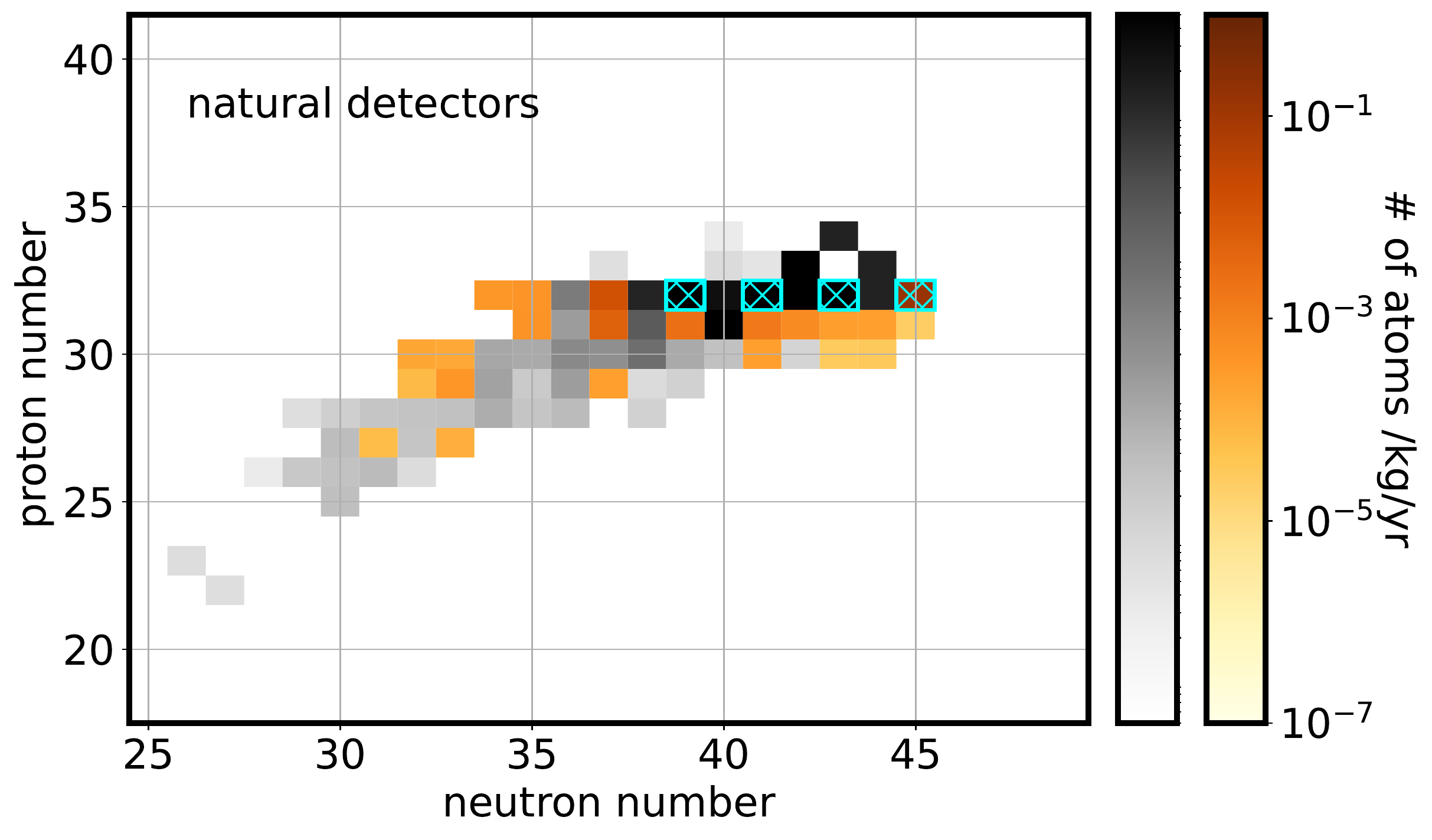}
\includegraphics[width=0.9\columnwidth,keepaspectratio=true]{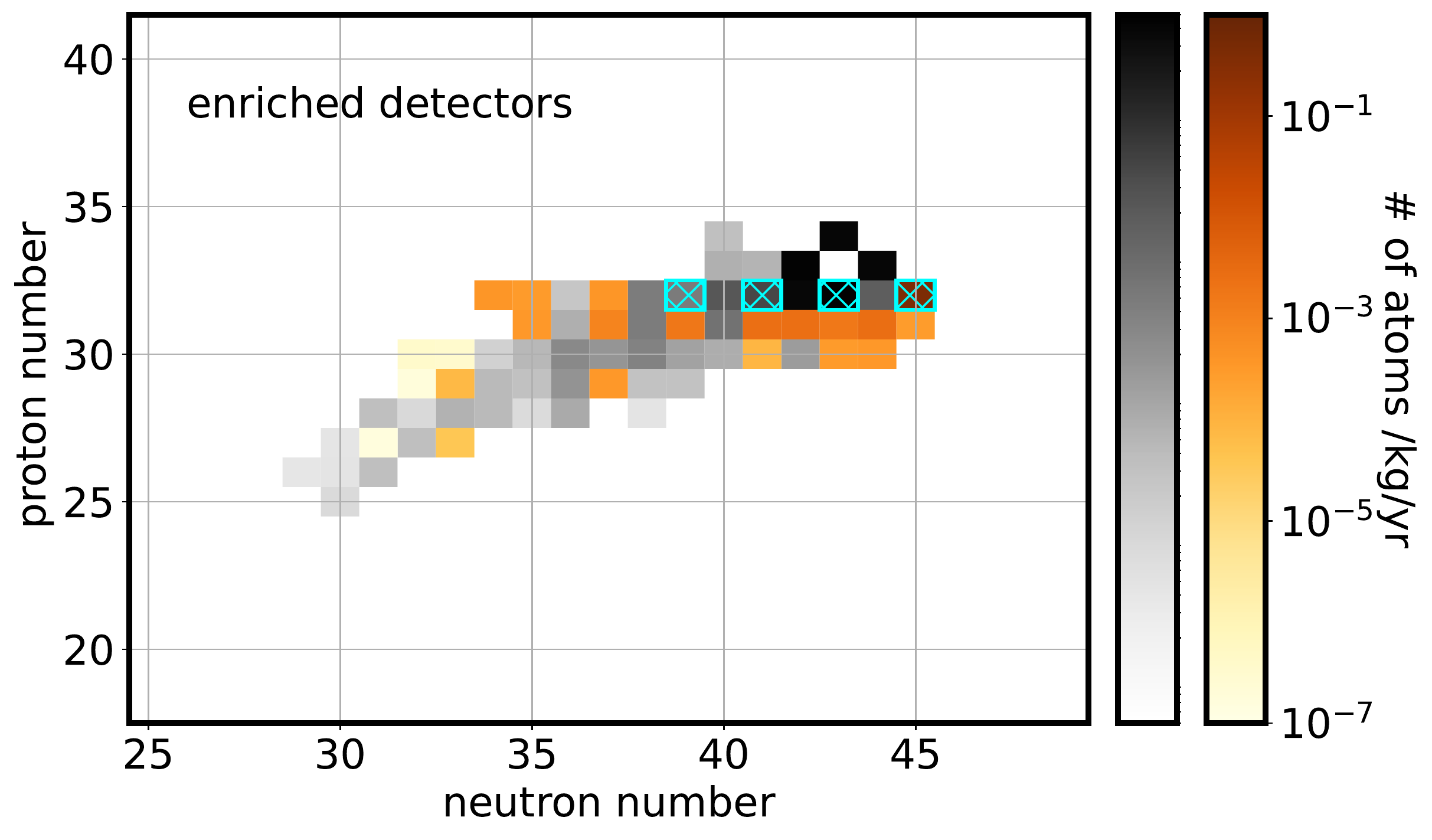}
\caption{(Color online) Production rate of isotopes from in-situ cosmogenics and their products with natural detectors (top) and enriched (87\% $^{76}$Ge) detectors (bottom). The colored scale represents isotopes with the potential to contribute background for \nonubb~while the grey-scale isotopes do not contribute to the region of interest (ROI). The germanium isotopes with odd neutron number analyzed in this paper are outlined in cyan.}
\label{fig:IsotopesProduced}
\end{figure}

In germanium, the production of neutron-induced isotopes has been studied with AmBe neutron sources~\cite{Mei2008} and neutron beams~\cite{Elliott2010}. It has been shown that a number of long-lived isotopes such as $^{57}$Co, $^{54}$Mn, $^{68}$Ge, $^{65}$Zn, and $^{60}$Co are produced~\cite{Miley1992,Avignone1992, Mei2009,Cebrian2010}. 
These isotopes, as well as others, are also generated when the germanium detectors are fabricated and transported at the surface. This is a well-known problem~\cite{Avignone1992,BARABANOV2006115}, and special precautions were taken in the production of \MJ~detector crystals~\cite{ABGRALL2018314}, including use of a database with detailed tracking of surface exposure~\cite{ABGRALL201552}. Once underground, the flux of cosmic rays is significantly reduced, but not zero. For double-beta decay searches in $^{76}$Ge, the isotope $^{68}$Ge is often considered as one of the major background contributors~\cite{Elliott2010,Domula:2014tnt}. It is created by spallation reactions on germanium by muons, or by fast neutrons energies of several tens of MeV. Its 271-day half-life renders it impossible to correlate the decay signal with the incident cosmogenic shower that produced it. Its radioactive daughter $^{68}$Ga (Q-value 2.9\,MeV) has a decay energy spectrum that spans over the region of interest (ROI) for \nonubb~in $^{76}$Ge (2.039 MeV). A number of other isotopes are produced in spallation reactions with muons, high-energy photons, or fast neutrons interacting with the nuclei. In addition to these, $^{77}$Ge can be produced via neutron capture reactions, which primarily occur at lower neutron energies. Figure~\ref{fig:IsotopesProduced} shows the production rate of isotopes created inside the germanium crystals during simulations of cosmogenic muons interacting with the \DEM, and the close-by rock. As shown and discussed later in detail, the isotopic composition of the germanium detectors will affect the rate of production of the isotopes.

In this paper, we report on the production rate of meta-stable states in the isotopes $^{71m}$Ge, $^{73m}$Ge, $^{75m}$Ge, and $^{77m/77}$Ge and compare to predictions from simulations. Given the ultra-low radioactive background of the \DEM, we can use specific signatures to identify these isomeric decays. Therefore, we analyze the pulse-shape of the signal waveform which occur after incoming muons. Similar experiments used the time between initial muon interaction and a subsequent decay, such as Borexino~\cite{Back2006, Bellini2011}, KamLAND~\cite{Abe2010}, Super-Kamiokande~\cite{Li2015,Zhang2016}, and SNO+~\cite{Aharmim:2019yng,Anderson:2020nym}. Incoming muon and their showers interact with these large experiments, and in-situ activation can be an important background. In current generation experiment, the background from cosmogenics and neutron-induced isotopes is not significant. However, its significance increases with the size and decreasing background goals of future generation efforts. In the following, we will describe the isotope signatures used as well as the search in the \DEM~data. This section is followed by a comparison to rates from simulations using \Geant4 and FLUKA. We conclude by discussing the estimated impact on the tonne-scale effort, the Large Enriched Germanium Experiment for Neutrinoless double-beta Decay (LEGEND)~\cite{LEGEND2021}.

\section{Search for in-situ activation signatures in the \MJD}
\label{sec:1}

\subsection{The \MJD}

\begin{figure}[t]
\includegraphics[width=0.95\columnwidth]{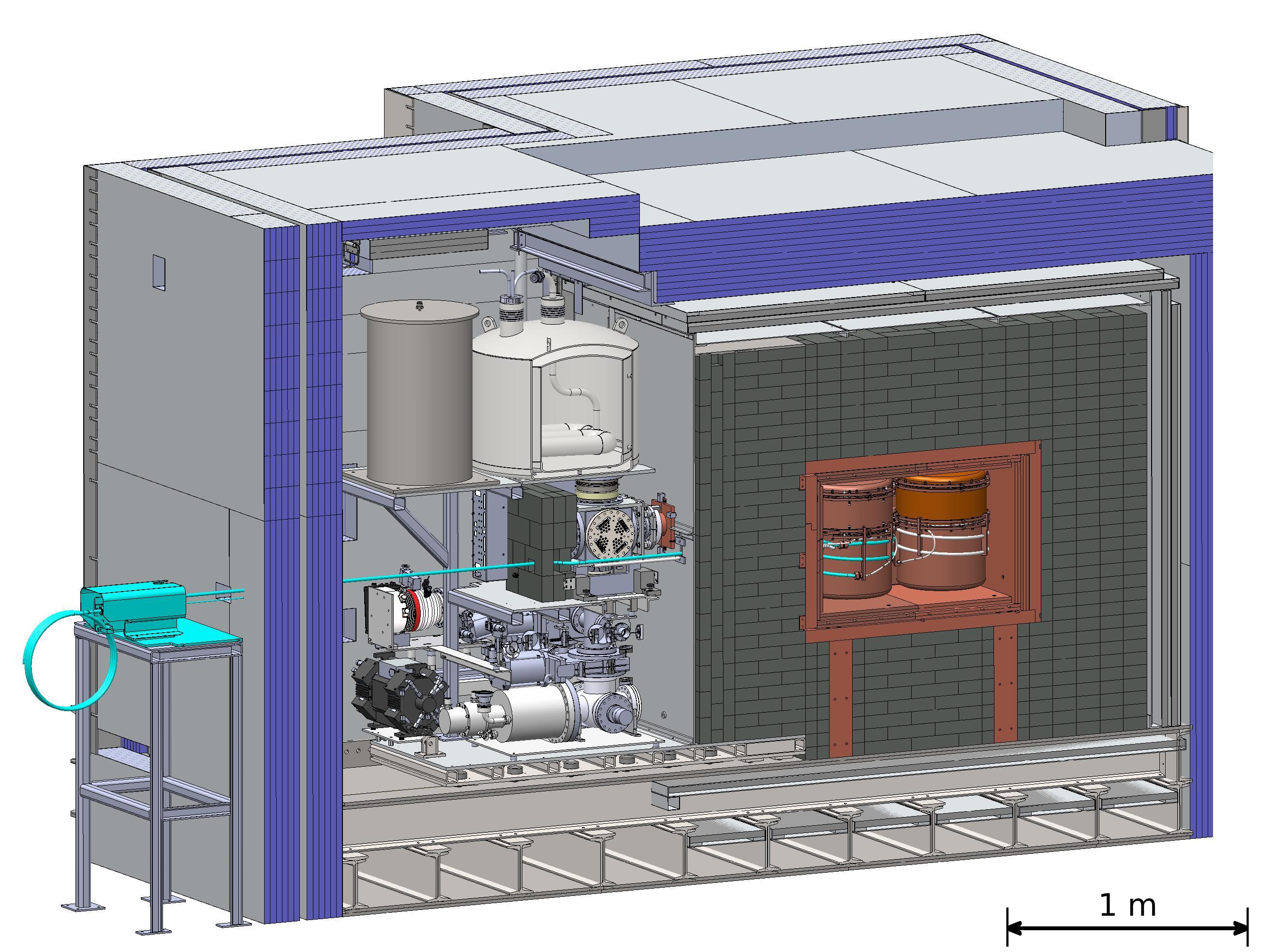}
\caption{(Color online) Cross-sectional drawing of \MJD\ including besides the detector cryostats also cryogenic systems, vacuum hardware, and shielding layers. Copper shielding is shown in brown, lead bricks in dark gray and the poly shield in purple. Not all muon veto panels are shown for better visibility.}
\label{fig:Demonstrator}
\end{figure}

The \MJD~contained fifty-eight p-type point contact (PPC) germanium detectors installed in two independent cryostats, totalling 44.1\,kg of high-purity germanium detectors. Of these, 29.7\,kg are enriched up to 87\% in $^{76}$Ge~\cite{Aalseth2018,ABGRALL2018314}, see Table~\ref{table:detisotopes}. Each germanium crystal was assembled into a detector unit and stacked in strings of three, four, or five units. Each cryostat contained 7 strings. The mass, diameter, and height of each crystal ranged from 0.5 to 1\,kg, 6 to 8\,cm, and 3 to 6.5\,cm, respectively. There were several shielding layers around the cryostats. From outside to inside these were: a 12-inch thick polyethylene wall, a muon veto made of plastic scintillator, a radon exclusion box purged with liquid nitrogen boil-off, an 18-inch thick lead shield, and an innermost a 4-inch thick copper shield, see Fig.~\ref{fig:Demonstrator}. The innermost cryostats and the inner structural material were made of ultra-pure, underground electroformed copper which contains extremely low levels of radioactivity from thorium and uranium~\cite{Abgrall2016}.

\begin{table*}
\begin{tabular}{c|c|c}
 \hline
 Isotope & Natural detector & Enriched detector \\
 & \% & \% \\
 \hline \hline
 $^{70}$Ge & 20.3 $\pm$ 0.2 & 0.004 $\pm$ 0.003 \\
 $^{72}$Ge & 27.3 $\pm$ 0.3 & 0.009 $\pm$ 0.004 \\
 $^{73}$Ge & 7.76 $\pm$ 0.08 & 0.028 $\pm$ 0.004 \\
 $^{74}$Ge & 36.7 $\pm$ 0.2 & 12.65 $\pm$ 0.14 \\
 $^{76}$Ge & 7.83 $\pm$ 0.07 & 87.31 $\pm$ 0.14 \\
 \hline \hline
\end{tabular}
\caption{Isotope composition of the \MJD's detectors}
\label{table:detisotopes}

\end{table*}
Data sets used in this analysis were acquired over the course of almost 4 years, from 2015 until 2019 --- the same data used in Ref.~\cite{Alvis_2019}, with a similar blinded analysis scheme. All analysis routines are fixed and reviewed on open data, before being applied to the full data set after unblinding. The total exposure for this analysis is $9.4\pm0.2$\,kg\,yr and $26.0\pm0.5$\,kg\,yr for the natural and enriched detectors, respectively~\cite{Alvis_2019}. The signals from each detector are split into two different amplification channels. The high-gain channels reach from a\,keV-scale threshold up to about 3\,MeV and allow an excellent pulse shape analysis for low-energy physics searches as well as double-beta decay analysis. The low-gain data spans up to 10-11\,MeV before saturating, allowing for searches and analyses of high-energy backgrounds. The decay pattern presented here are in the energy range of tens of keV up to MeV. Detector signals include waveforms with duration 20\,$\mu$s followed by a dead time of 62~$\mu$s. Some portion of the data used multi-sampling of waveforms which extended length allowed better pulse-shape analysis in the \nonubb\ analysis, see Ref.\,\cite{Alvis_2019}, with a duration of 38.2\,$\mu$s and a dead time of 100~$\mu$s. The rising edge is located at a timestamp of $\sim$10~$\mu$s from the beginning of the waveform. Given a distinctive waveform structure and short time-delayed coincidence, the searches for $^{73m}$Ge and $^{77}$Ge are almost background-free. By taking advantage of the low count-rate and excellent energy resolution of the \DEM, the production rate of $^{71m}$Ge, $^{75m}$Ge and $^{77m}$Ge can also be determined.

\subsection{Search for $^{73m}$Ge}
\label{sec:73Ge}
\begin{figure}[t]
\includegraphics[width=0.5\textwidth]{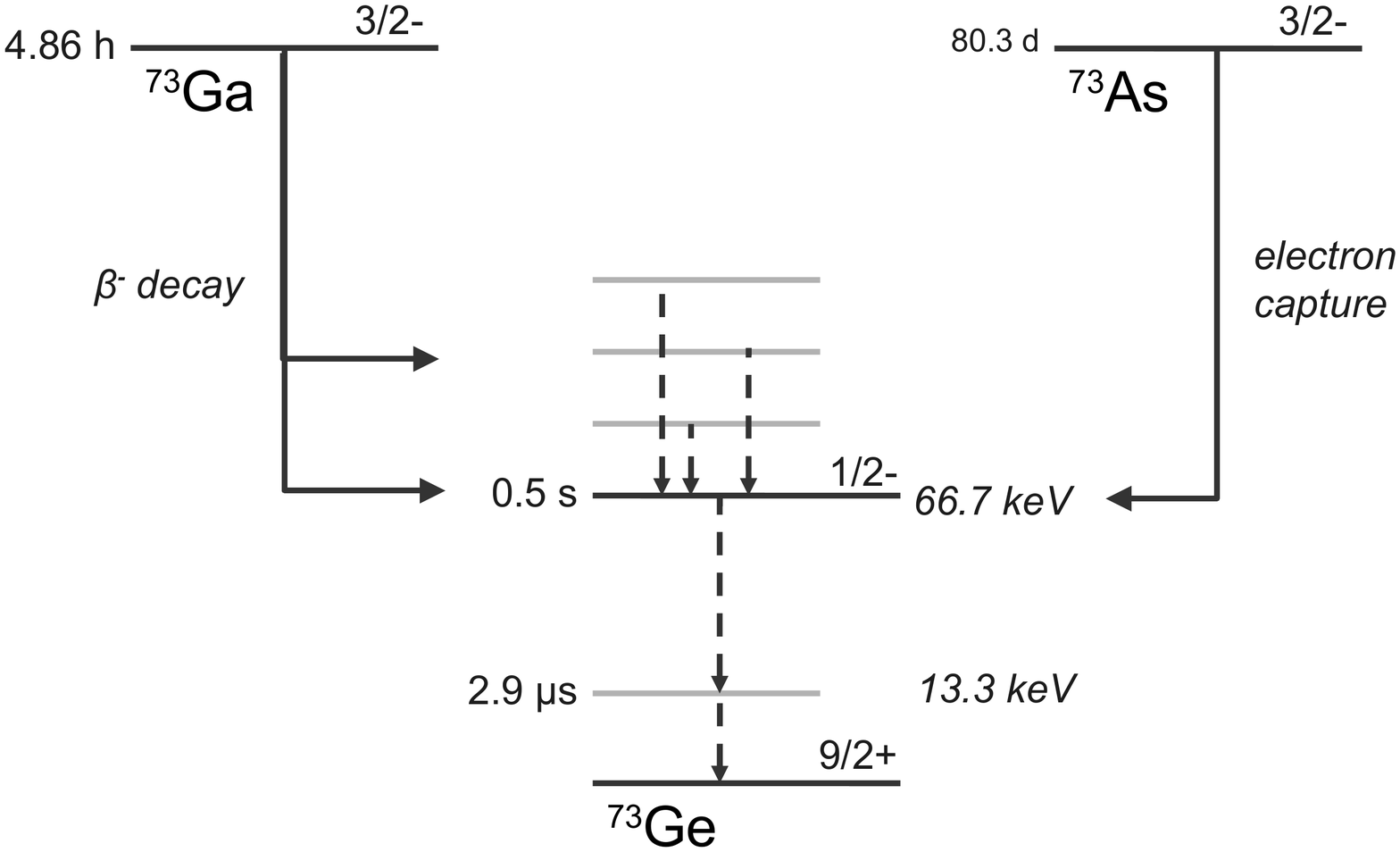}
\caption{The decay scheme of $^{73}$Ga, $^{73m}$Ge, and $^{73}$As to $^{73}$Ge ~\cite{Firestone1999, NNDC}.}
\label{fig:73GaTo73Ge}
\end{figure}

\begin{figure}[htb]
\centering
\includegraphics[width=0.95\columnwidth, keepaspectratio=true ]{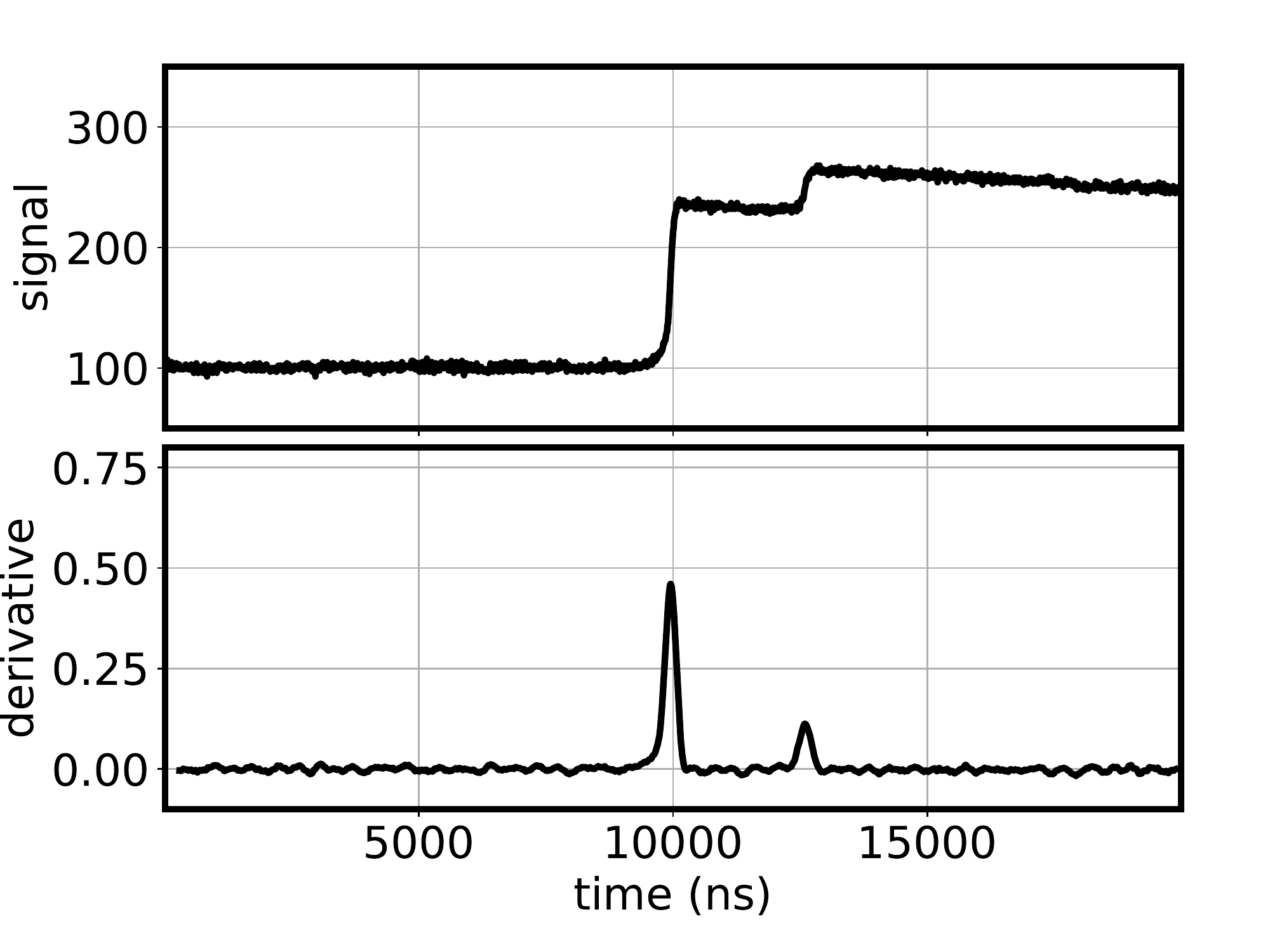}
\caption{Top: Two-step waveform ($second$ event); Bottom: The first derivative (current) of the waveform. A clear two-step pattern can be observed due to the 53\,keV and 13\,keV transitions in sequence.}
\label{fig:waveform_18672_51097_1298}
\end{figure}

One can consider both of the first two excited states in $^{73}$Ge to be isomers since their half-lives are longer than usual for nuclear states. The second excited state has a half-life $T_{1/2}$ of about 0.5\,seconds and is named $^{73m}$Ge within this work. Most $\beta$-decays from neighboring isotopes populate this state as shown in Fig.~\ref{fig:73GaTo73Ge}. In addition, de-excitations from higher excited states within $^{73}$Ge can feed this state, due to inelastic scattering of neutrons, photons, or other particles. The half-life of $^{73m}$Ge is long enough to apply a time-delayed coincidence method~\cite{DeBenedetti1946, Bashandy1959}. After an energy deposition by an initial decay or de-excitation (first event), a second event can be observed. The second event is the de-excitation of the meta-stable state at 66.7\,keV. The analysis aims to identify two events in one detector within a short time window, with the second event possessing a specific energy and structure. The individual detector count-rate is about $10^{-4}$\,Hz over the entire energy spectrum. The probability for a second event in a 5-second long window (10$\times T_{1/2}$) is less than 0.05\% for any two random events. After applying the energy requirement on the second event, the search becomes quasi background-free. The de-excitation of the 66.7-keV state can be identified uniquely since it is a two-step transition, as seen in Fig.~\ref{fig:waveform_18672_51097_1298}. First, an energy of 53.4\,keV is released when relaxing to the first excited state. It is followed by a 13.3-keV pulse that has a half-life of 2.95\,$\mu$s. This is short enough to be observed within a single waveform and has a distinctive pattern. 


The data is first scanned with a simple energy acceptance window using the \MJ~standard energy calibration~\cite{Alvis_2019}. When the two transitions (53 and 13\,keV) are well separated in time, the energy of the event is flagged in the data as the energy of the first transition around 53\,keV. If the two transitions are very close in time and look like a single waveform, the energy appears as the sum of the two steps. Potential background like in-detector Compton scattering would also show such very short step structure, and are suppressed by the later requirements. Including the energy resolution of about 0.5\,keV at these energies, this first algorithm creates a selection of candidates between 48 and 72\,keV with negligible efficiency loss. For each of these $second$ event candidates, the preceding five seconds of data is scanned for a possible $first$ event. All events above the general analysis threshold of 5\,keV is accepted, and only clearly identified noise bursts~\cite{Alvis:2019dzt} are rejected. Only delayed coincidence combinations that fulfill these basic conditions are fed into the detailed analysis searching for the two-step pattern, since this part of the analysis is computationally intense. 

For the $^{73m}$Ge decay search, a special pulse shape analysis is applied to identify the short-time delayed-coincidence waveforms. As shown in Fig.~\ref{fig:waveform_18672_51097_1298}, a clear two-peak pattern in the first derivative of the waveform can be found. The amplitude ratio of the two peaks is roughly equivalent to the energy ratio of the two transitions (53/13$\approx$4). The delay between the two peaks is comparable to the lifetime of the first excited state($\sim3\mu\text{s}$). Noise and slow waveforms~\cite{PhysRevLett.118.161801} are rejected by requiring narrow peaks. 
To estimate the background of the analysis, we removed the need for a $first$ event, and repeated the analysis. Over the whole data set, three pile-up events were found within the same energy window and the correct ratio between the two signals but outside the delayed-coincidence time window. These can be interpreted as random coincidences with a rate of 0.18\,cts/kg/yr. When combining this rate with the overall detector of 10$^{-4}$\,Hz, we assume this background negligible for the further analysis. Since two-step waveforms of the appropriate energy and peak ratios are rare, the analysis efficiencies were estimated using simulated waveforms generated in germanium crystals by {\tt mj\_siggen}~\cite{Radford:2014}. A two-step waveform can be formed by combining one 53-keV waveform and one 13-keV waveform with a short-time delay determined in accordance with the half-life 3\,$\mu$s. The acceptance windows of the simulation analysis parameters were set conservatively in a $\pm$3$\sigma$ range. The uncertainty of the analysis cuts was estimated with two-step waveforms generated by combining 53\,keV waveforms and 13\,keV waveforms from calibration data \cite{ABGRALL201716}. Negligible differences between simulated waveforms and combined calibration waveforms were found. These differences can be attributed to the additional baseline noise of the second waveform, as well as the existence of a small population of slow waveforms in the calibration data. While the initial energy acceptance and time search has only minimal efficiency loss, the waveform analysis is not 100\% efficient because of the length of the recorded waveform and the efficiency to distinguish the two-step pattern. The final combined efficiency of the analysis chain $\epsilon_{tot} =$ is $79\pm14\%$ for normal sampling and $88\pm14\%$ for data sets taken with multi-sampling.

\begin{table*}
\begin{tabular}{c|c|c|c|c|c|c|c}
 \hline
 Event & Energy of the first event & $\Delta T_1$ & $\Delta T_2$ & $\Delta T_{\mu} $ & Ratio & Enriched & Time underground\\
 & (keV) & (s) & ($\mu$s) & (s) & $E_1/E_2$ & detector & (Date$_\text{UG}$ : Date$_\text{Event}$ : $\Delta T_\text{UG}$ (months))\\
 \hline \hline
 1 & 2864.3 & 0.5 & 1.2 & 168.2 & 4.1 & No & 11/2010 : 09/2015 : 59 \\
 \hline
 2 & 325.8 & 0.1 & 0.8 & 5930.2 & 4.0 & No & 11/2010 : 09/2015 : 59\\
 & 738.7 & & & & & \\
 \hline
 3& 157.1 & 0.3 & 2.7 & 0.3 & 4.0 & No & 11/2010 : 09/2016 : 71\\
 & 308.0 & & & & & \\
 & 7.8 & & && &\\
 \hline 
 4$^*$ & 10.9 & 0.2 & 2.6 & 2128.9 & 4.1 & Yes & 06/2015 : 10/2016 : 16 \\
 \hline
 5$^*$ & 11.2 & 0.6 & 6.2 & 2314.3 & 3.9& Yes & 08/2015 : 11/2016 : 15 \\
 \hline
 6$^*$ & 11.0 & 2.5 & 3.8 & 462.3 & 4.2 & Yes & 07/2015 : 03/2017 : 20\\
 \hline
 7 & 883.6 & 1.0 & 1.1 & 1029.7 & 3.7 & Yes & 01/2013 : 03/2018 : 63\\
 \hline
 \hline
\end{tabular}
\caption{The candidates of $^{73m}$Ge decays that pass all analysis steps. Two or more energies for the first events indicate events for which more than one detector was triggered, as could be the case when a neutron scatters. The energy of the second event is not listed, since it is restricted as described in the text. $\Delta T_1$ is the time difference between the $first$ and $second$ events. $\Delta T_2$ is the time difference of the two steps in the $second$ event waveform. The time relative to the last muon identified by the muon veto is given as $\Delta T_\mu$. The ratio $E_1 / E_2 $ indicates the amplitude ratio of the two peaks in the first derivative of the short time-delayed coincidence waveform of the $second$ event. ``Enriched Detector" indicates whether or not the event occurred in an enriched detector. Events marked with $^*$ are considered background from surface activation due to their energy and distribution. The last column represents the date that the detector went underground (Date$_\text{UG}$), the month the event occurred in the data stream (Date$_\text{Event}$), and the time spent underground ($\Delta T_\text{UG}$).}
\label{table:data_73Ga}
\end{table*}

\begin{figure}[t]
\includegraphics[width=0.5\textwidth]{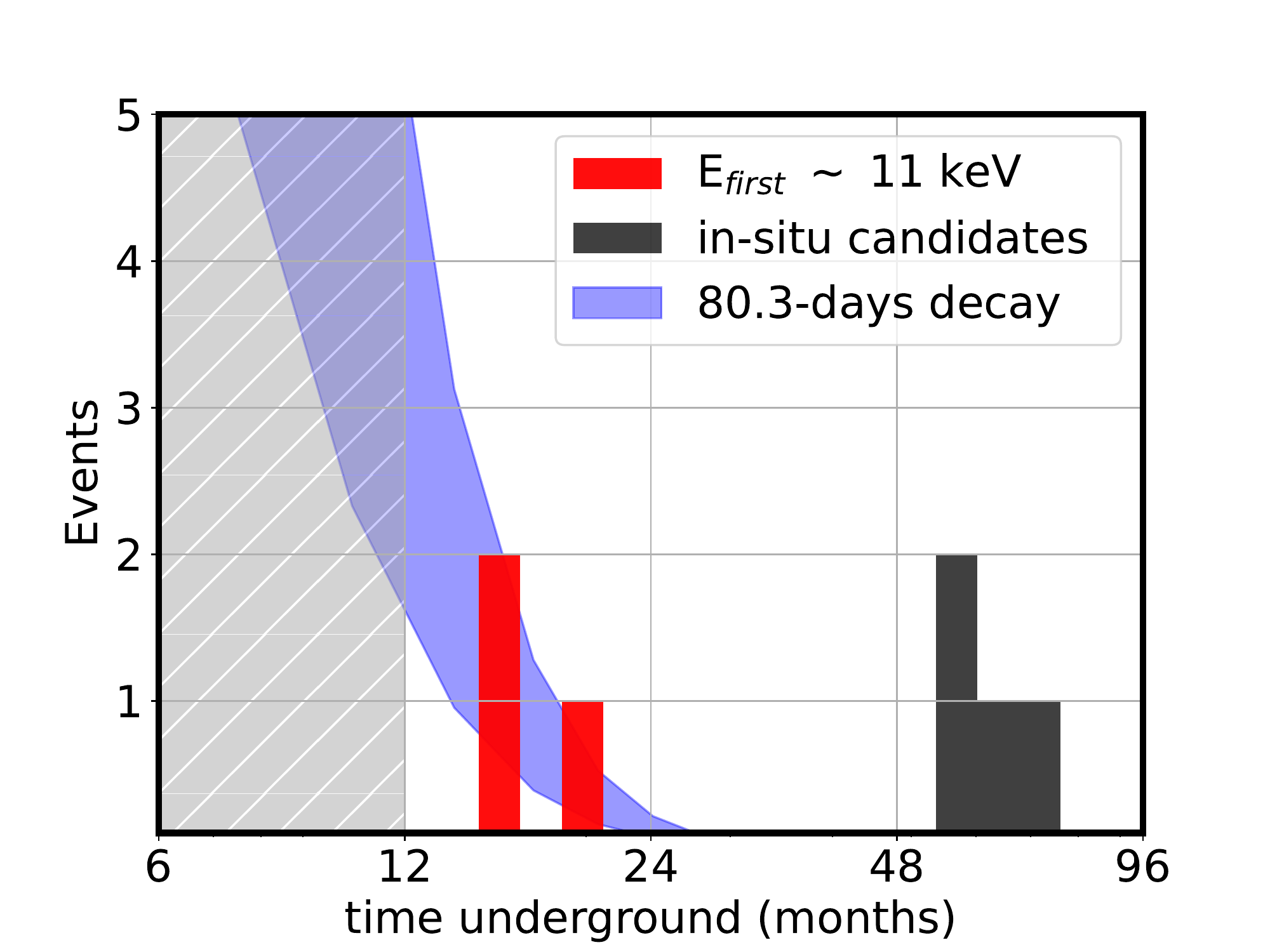}
\caption{(color online) The distribution of $^{73m}$Ge candidate events as a function of the time spent underground. Events that are considered of $^{73}$As origin due to their 11\,keV x-ray signature are shown in red, together with a fitted decay curve using an 80.3-day half-life (blue band). Based on the three arsenic events, this curve shows the scale of the $^{73}$As background within $^{73m}$Ge search over time. All other events are shown in black. The grey area indicates the time before data taking.}
\label{fig:73As_LifeTime}
\end{figure}

Table~\ref{table:data_73Ga} shows the list of $^{73m}$Ge candidates identified. Three of the candidates show a first event with energy around 11\,keV. These events are likely due to a $^{73}$As electron capture decay ($T_{1/2}$ = 80.3 days), cf. Fig.~\ref{fig:73GaTo73Ge}. The isotope $^{73}$As can be cosmogenically generated on the surface before detectors arrive underground. The cool-down time between the day detectors arrive at the 4850-foot level and start of data taking differs from detector to detector, from about a year to several years. All arsenic-type events occurred in the last batch of detectors brought underground, see Fig.~\ref{fig:73As_LifeTime}. Detectors which were brought underground earlier have no such signature observed, supporting this assumption. Simulations predict that only a negligible amount of $^{73}$As was produced in-situ. Therefore, we excluded these three events from our cosmogenic analysis. The identification of these events illustrates the high sensitivity of the $^{73m}$Ge tagging process. The remaining events are used to determine the isotope production rate. The statistical uncertainty for a 1-$\sigma$ confidence level is determined using the Feldman-Cousins approach~\cite{Feldman1998}. The systematic effects due to the analysis procedure are on the order of 14$\%$. These uncertainties include effects like dead-time windows after a trigger, as well as periods in which a selection of events was not possible, e.g. when transitioning to a calibration. The final isotope production rate is $0.38^{+0.34}_{-0.19}$ and $0.05^{+0.09}_{-0.02}$ cts/(kg\,yr) for the natural and enriched detectors, respectively. A comparison with simulation is shown in Table~\ref{table:ExpSim}.

\subsection{Search for $^{77}$Ge}
\begin{figure}[t]
\includegraphics[width=0.55\textwidth]{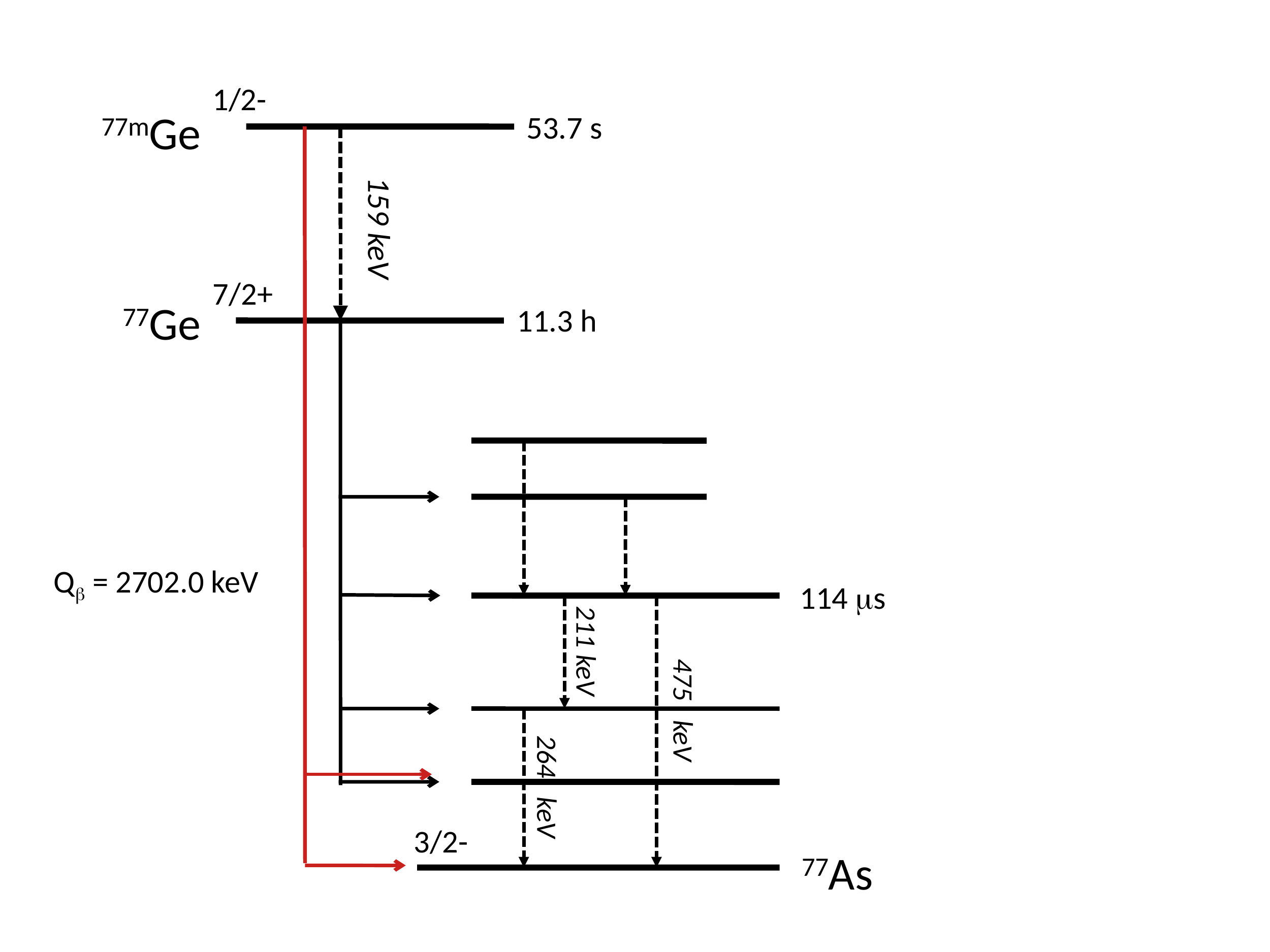}
\caption{The decay scheme of $^{77}$Ge and $^{77m}$Ge (red) to $^{77}$As~\cite{Firestone1999, NNDC}.}
\label{fig:77GeTo77As}
\end{figure}

The isotope $^{77}$Ge is produced by neutron capture on $^{76}$Ge. After the capture, the excited nucleus decays either to the ground state of $^{77}$Ge or to the meta-stable state at 159\,keV ($^{77m}$Ge). The neutron capture cross-section for each has been measured~\cite{Bhike2015}. Both states can decay to $^{77}$As with distinct half-lives and gamma emissions, cf. Fig.~\ref{fig:77GeTo77As}. The $^{77m}$Ge decay can release up to 2.86\,MeV in energy. In more than half of the decays the final state of the $\beta$-decay is the ground state of $^{77}$As. In these cases, the single $\beta$ particle can produce a point-like energy deposition similar to that of neutrinoless double-beta decay. Its relatively short half-life of only 52.9 seconds allows for the introduction of a time-delayed coincidence cut as suggested by Ref.~\cite{Wiesinger_2018}. The decay of $^{77}$Ge also spans over the \nonubb~ROI. However, the populated higher-energetic states of $^{77}$As will decay via gamma emission. This additional photon allows a background-suppression by analysis cuts such as multi-site event discrimination~\cite{Alvis:2019dzt}, multi-detector signatures, or an argon veto anti-coincidence~\cite{Wiesinger_2018}. For this study, we can use the 475\,keV state of $^{77}$As and its half-life of 114 $\mu$s to identify the creation of $^{77}$Ge. Similar to the search for $^{73m}$Ge, the time-delayed coincidence method is used. A $first$ event from the $\beta$-decay of $^{77}$Ge is followed by a $second$ event with a well-defined energy of 475\,keV. Also included in the analysis is the search for the branch that includes a 211 or 264\,keV transition, as shown in Fig.~\ref{fig:77GeTo77As}. Since the half-life of the meta-stable state in $^{77}$As is shorter than in the $^{73}$Ge case, the de-excitation to the ground state has a significant chance to occur in the dead time period of the previous $first$ decay event. Therefore, the detection efficiency compared to the $^{73m}$Ge search is reduced to 69$\%$ (54$\%$) for normal (multi-sampled) waveforms. Full energy detection efficiency of about 54\% for these $\gamma$ rays was estimated with the \MaGe~simulation code~\cite{Boswell2011}. The total efficiency includes branching effects in the decay scheme and is calculated to be 31$\%$ (25$\%$) for normal (multi-sampled) waveforms. Due to the extremely low total event rate in each detector of about $10^{-4}$Hz, the number of expected background events is on the order of $10^{-7}$ for the whole data set. No candidate event was found in the current search. The Feldman-Cousins method was used to estimate the uncertainty with the assumption of zero background. Since no events were found, an upper limit on the event rate can be set to less than 0.7 and 0.3 cts/(kg\,yr) for the natural and enriched detectors, respectively. 

\subsection{Search for $^{71m}$Ge, $^{75m}$Ge, and $^{77m}$Ge }

\begin{figure}[t]
\includegraphics[width=0.9\columnwidth]{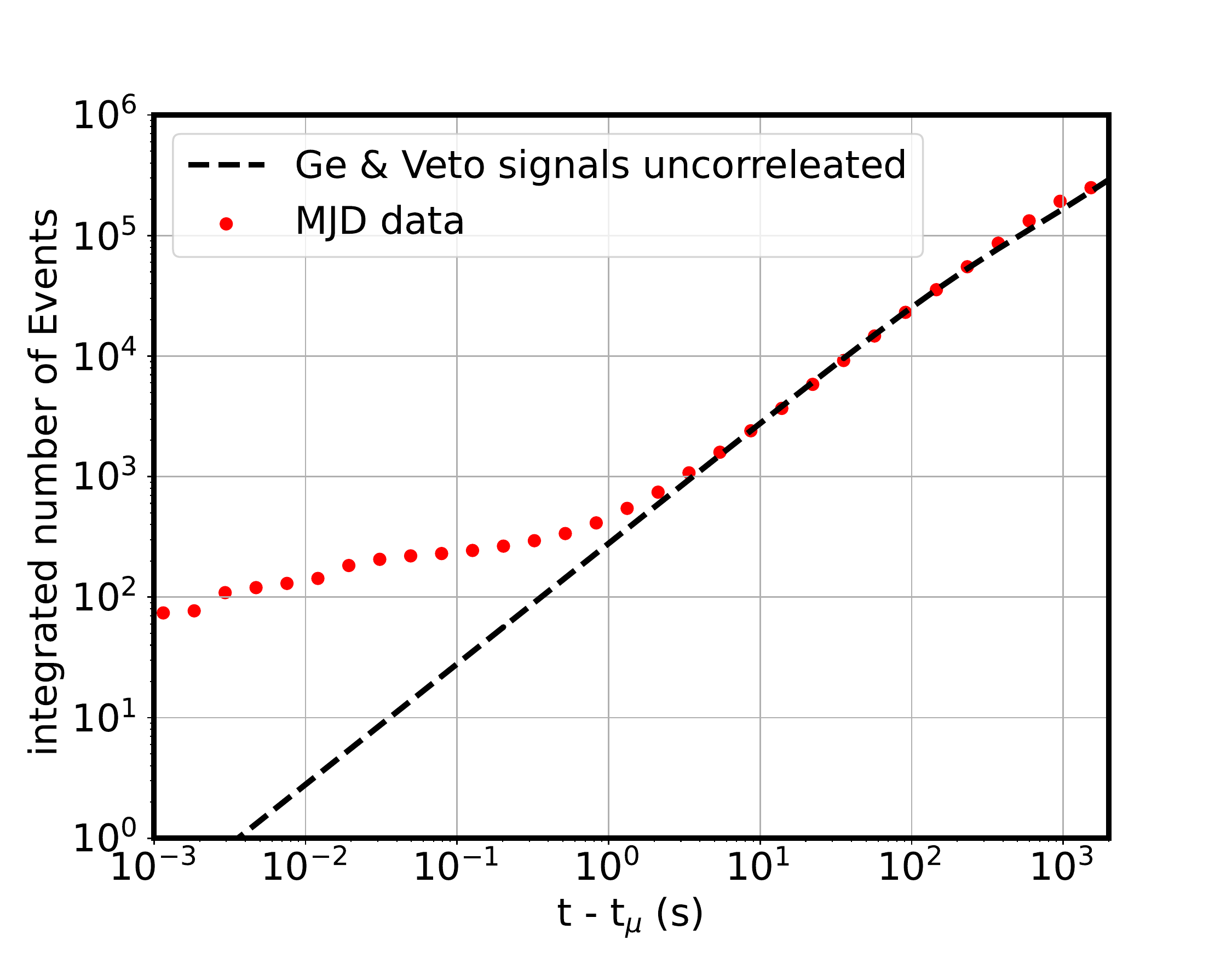}
\caption{(Color online) The red dotted curve shows the integrated number of events above the analysis energy threshold between a time $t$ and the previous muon at time $t_\mu$ in the \DEM~data. The black dashed line represents the expected number of events calculated assuming that the rates for the muon system and germanium array would be completely independent. For long times, the trend corresponds to a random coincidence; however, for short time windows a deviation from the independent random triggering can be found which illustrates that there is a clear correlated contribution by muons in both systems.}
\label{fig:DTmuons}
\end{figure}

For many germanium isotopes with odd neutron number, low-lying isomeric states exist. The half-lives of these states range from a few ms for $^{71m}$Ge to almost a minute for $^{77m}$Ge. When muons and their showers pass through the \DEM, they can cause knock-out reactions on the stable germanium isotopes. These reactions, dominated by neutrons or photons, create excited odd-numbered germanium isotopes, which populate these isomeric states when relaxing. When decaying, each isomer has a characteristic energy release of a few hundred\,keV. This delayed energy release, in combination with the \DEM's low count rate, enables a search for signatures from these isotopes. A first event is identified as a muon using the scintillator-based muon veto system as described in Ref.~\cite{Abgrall2017}. Second events are searched for after the timestamp of the muon event in the germanium data stream. These second events have a characteristic transition energy from the isomeric state to the ground state, see Table~\ref{table:data_isomers}. The energy windows of the event selection are $\pm$5\,keV around the expected energy and the time windows are five to ten times the corresponding isomer half-lives after the incident muon. The uncertainty of the veto-germanium timing is known to be negligible relative to the time considered. Efficiency values to detect signatures based on \MaGe\ for each of the corresponding signatures are given in Table~\ref{table:data_isomers}. 
To estimate the rate of random background for each signature, we considered the overall signal rate and the muon flux. In a germanium detector, the overall signal rate is about 0.05-0.2 events per day per detector in a 10\,keV wide window for the energies of interest~\cite{Aalseth2018}. The muon flux at the 4850-ft level~\cite{Abgrall2017} is measured to be about 6 muons per day passing through the experimental apparatus. The overlap of both distributions can be used to estimate the background rate at the expected transition energy and time window (see Table~\ref{table:data_isomers}). While the time windows of $^{75m}$Ge and $^{77m}$Ge are about 5 times longer than their half-lives, the time window of $^{71m}$Ge is chosen to be 10 times the half-life. This was done to decrease the effect of statistical fluctuations that can be present in short time windows when estimating the background. 
The number of events based on these two rates as a function of time between muon and germanium events was calculated to verify this estimate. Figure~\ref{fig:DTmuons} shows the time of events in the \DEM 's germanium detectors relative to the time of the last muon compared to how the distribution would look like if the veto and germanium system would be not correlated. The number of events fits very well to the expected coincidental rate when the previous muon was more than one second before the germanium event. Additional events within one second of a muon are found and indicate a clear contribution from the muon-induced prompt backgrounds. Therefore, we give for rates that are consistent with upper limits, and for the $^{71m}$Ge-channel a rate over background, see Table~\ref{table:data_isomers}. These rates, combined with the rate of expected $^{73m}$Ge and $^{77}$Ge events, are now used to discuss the quality of simulations.

\begin{table*}
 \begin{tabular}{c|c|c|c|c|c|c}
 \hline
 Isotope & Transition energy & Half-life & Detection efficiency & Background estimate & Events found & Rate (UL) \\
 &  & & & nat/enr & nat/enr & nat/enr \\
 & (keV) & & ($\%$) & (cts) & (cts) & (cts/(kg yr)) \\
 \hline \hline
 $^{71m}$Ge & 198.4 & 20.4 ms & 67(5) & 0.13(1) / 0.29(3) & 4 / 6 & 0.6(4) / 0.3(2)\\
 \hline
 $^{75m}$Ge & 139.7 & 47.7 s & 91(5) & 99(14) / 189(20) & 104 / 213 & $<$1.9(1) / $<$1.7(1)\\
 \hline
 $^{77m}$Ge & 159.7 & 53.7 s & 15(1) & 82(13) / 194(21) & 81 / 194 & $<$6.4(4) / $<$5.8(3)\\
 \hline
 \hline
	\end{tabular}
\caption{Overview on the signatures of isomeric transition in odd germanium isotopes. The efficiency to detect these events includes the reduction due to branching in the decay. If the number of events is consistent with the background, upper limit calculations with 1$\sigma$ C.L. are given. The uncertainties for the individual rates are estimated in Table~\ref{table:ExpSim}. The efficiency of $^{77m}$Ge is reduced due to its high $\beta$-decay branching.}
\label{table:data_isomers}
\end{table*}

\section{Simulation of cosmogenic background in the \MJD}

\MaGe~\cite{Boswell2011} is a \Geant4-based \cite{Agostinelli2003} framework developed by the \MJ~and \Gerda~collaborations. The calculations were done with two different versions of \Geant, 4.9.6 and 4.10.5, with the same geometries to evaluate the consistency of the results. The first version coincided with the \DEM\ construction, while the latter was the version at the end of the data sets analyzed for this manuscript. This selection is arbitrary and newer versions are published more than once a year. Given the time-intense simulations, we restricted ourselves to these two versions in order to illustrate how results can change within one package, as discussed in Ref.\cite{ALLISON2016186}. In each case the physics list $\text{QGSP}\_\text{BIC}\_\text{HP}$ was used for simulations. This list uses ENDF/B-VII.1 data~\cite{Kahler2011, Chadwick2011} for nuclear reaction cross-sections and extrapolates into unmeasured energy regions or isotopes with TENDL~\cite{Koning2012}, a TALYS based evaluation ~\cite{TALYS2018}. In addition to the~\MaGe~based simulations, a simplified geometry was translated to FLUKA \cite{Bohlen2014}, version 2011 2x.6. Similar simulations were performed and the predicted isotope production rates were then compared to the \Geant4 output.

The muon flux at the Davis campus has been simulated~\cite{Abgrall2017} and was in good agreement with the measured values when the same distribution was used as the input. To study the results from each of the simulation packages, muons were generated inside a rock barrier surrounding the experimental cavity to allow the formation of showers. About four meters of rock are needed to fully develop all shower components ~\cite{Kluck2015}. Ten million muons were started as primaries on a surface above the~\DEM, equivalent to almost 200 years of measurement time. Two different geometries were used in the simulation. The first geometry is the early experimental configuration, representing about a year of \DEM~data where only half of the poly-shield was installed. In the second geometry, all of the 12-inch thick poly-shield was installed for the final configuration of the \DEM. Each simulated data set was weighted according to the exposure for each configuration, as given in Ref.~\cite{Alvis_2019}, and each data set reflects subsets of active and inactive detectors, respectively. 

\subsubsection{Isotope production rates}

In order to understand which isotopes are produced, the rate of each isotope created by muon interactions in the \DEM~is calculated from the simulation. As shown in Fig.~\ref{fig:IsotopesProduced} the difference in isotopic mixtures creates a wide variety of isotopes. Isotopes that are created in spallation reactions can create daughter isotopes during the subsequent $\beta$-decays and electron captures. A natural isotope mixture in germanium tends to produce lighter isotopes than the enriched mixture. In the \DEM's enriched material, fewer isotopes with neutron numbers less than 42 can be found because spallation reactions have to knock out additional nucleons to produce these. The rates for these higher energy spallation reactions are suppressed because of the decreased flux of higher energy projectiles, as well as smaller reaction cross-sections.

\begin{figure}[t]
\centering
\includegraphics[width=0.9\columnwidth,keepaspectratio=true]{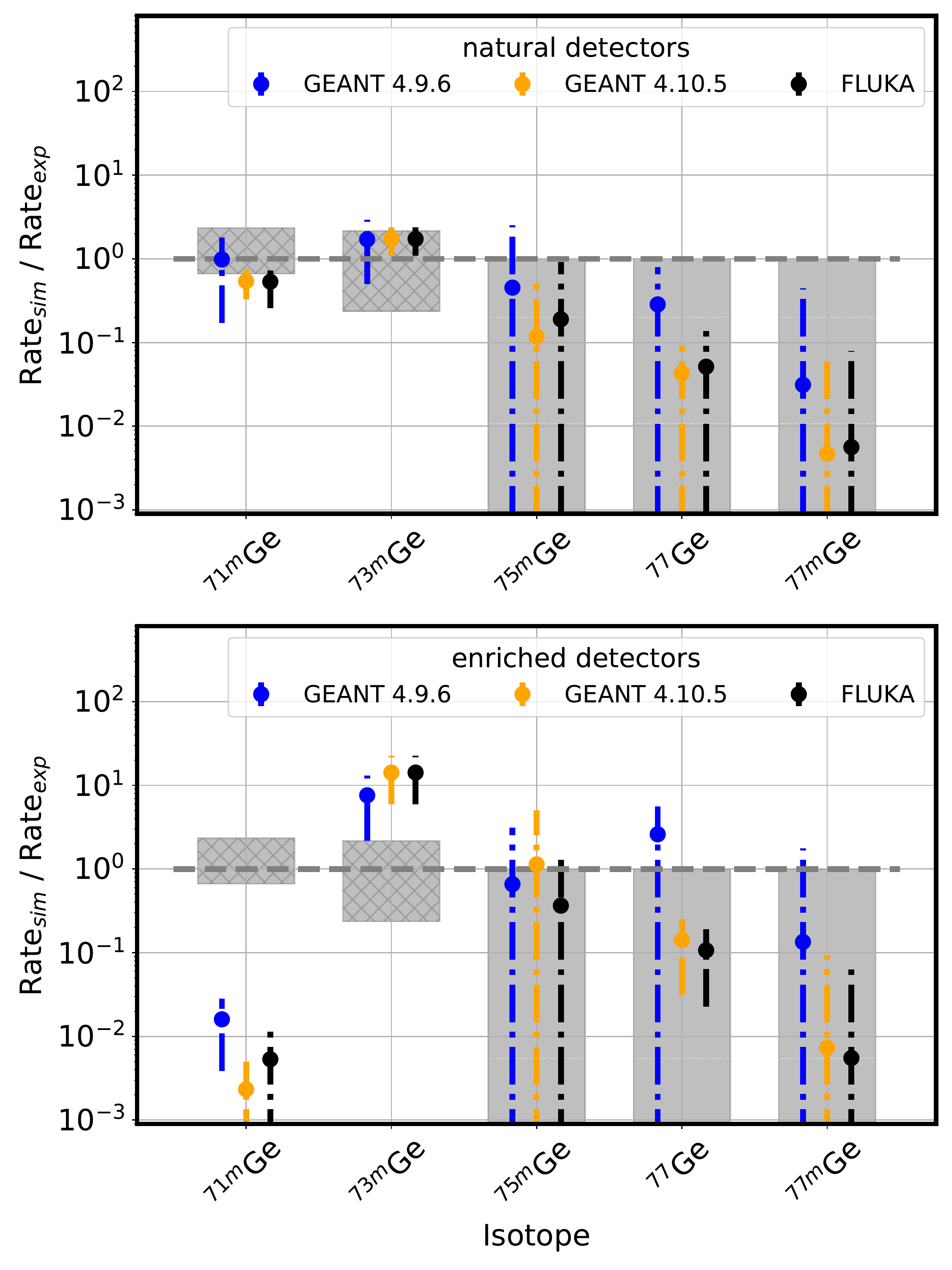}
\caption{(Color online) Comparison of each simulated rate relative to the experimental rate as given in Table~\ref{table:ExpSim} for natural Ge (top) and the \MJ~enriched Ge (bottom). A ratio of one would indicate that the simulation is in good agreement with the experimental findings. If no counts were observed, the expected upper limit was used as the experimental rate. The grey shaded areas show the uncertainties based on the experimental rate; the error bars on the data points represent the uncertainties in the simulations. }
\label{fig:SimCodes}
\end{figure}

A comparison of the three simulations with the experimental data can be found in Table~\ref{table:ExpSim}. When neutron capture occurs on $^{76}$Ge, \Geant4 populates the ground state $^{77}$Ge exclusively. Using the cross-sections in Ref.~\cite{Bhike2015}, an expected production rate of $^{77m}$Ge was calculated based on the rate of ground-state production, and the meta-stable isotopes were then added to the simulation manually, a method similar to Ref.~\cite{Wiesinger_2018}. For spallation reactions, isomeric states are created, so no correction was necessary. While the overall agreement is good, none of the simulation packages is able to reproduce all the experimental rates, as seen in Fig.~\ref{fig:SimCodes}. Averaging the ratios between simulations and experiment for all isotopes considered, the simulations tend to overestimate production rates. However, this average is driven by the $^{73}$Ge ratio. Since the experimental rates have large statistical uncertainties, this trend might balance out.

\begin{table*}
 \begin{tabular}{c|c|c|c|c|c |c}
 \hline
 Isotope & Dominant production & Candidates & Experimental rate & \multicolumn{3}{c}{Simulated rate} \\
 & mechanism & & (cts/(kg yr)) & \multicolumn{3}{c}{(cts/(kg yr))} \\
 & & & & \Geant\ 4.9.6 & \Geant\ 4.10.5 & FLUKA \\
 \hline \hline
 & & \multicolumn{5}{c}{ } \\
 & & \multicolumn{5}{c}{natural detectors} \\
 & & \multicolumn{5}{c}{ } \\
 \hline\hline
  & & & & & & \\
 $^{71m}$Ge & $^{70}$Ge$(n,\gamma)$ & $4^{+2.8}_{-1.7}$ & $0.6^{+0.4}_{-0.2}$ & $0.59\pm0.33$ & $0.32\pm0.10$ & $0.32\pm0.08$ \\
 & & & & & & \\
 $^{73m}$Ge & $^{73}$Ge$(n,n')$, $^{74}$Ge$(n,2n)$ &$3^{+2.7}_{-1.5}$ & $0.38^{+0.34}_{-0.19}$ & $0.65\pm0.25$ & $0.63\pm0.16$ & $0.66\pm0.16$ \\
 & & & & & & \\
 $^{75m}$Ge & $^{74}$Ge$(n,\gamma)$ & $0^{+16}_{-0}$ & $0^{+1.9}_{-0}$ & $0.43\pm0.33$ & $0.11\pm0.03$ & $0.18\pm0.05$ \\
 & & & & & & \\
 $^{77}$Ge & $^{76}$Ge$(n,\gamma)$ & $0^{+1.3}_{-0.0}$ & $0^{+0.7}_{-0.0}$ & $0.10\pm0.04$ & $0.015\pm0.005$ & $0.026\pm0.011$ \\
 & & & & & & \\
 $^{77m}$Ge & $^{76}$Ge$(n,\gamma)$ & $0^{+9}_{-0}$ & $0^{+6.4}_{-0.0}$ & $0.10\pm0.04$ & $0.015\pm0.005$ & $0.018\pm0.009$\\ 
 & & & & & & \\
 \hline\hline
 & & \multicolumn{5}{c}{ } \\
 & & \multicolumn{5}{c}{enriched detectors} \\
 & & \multicolumn{5}{c}{ } \\ \hline\hline
 & & & & & & \\
 $^{71m}$Ge & $^{76}$Ge$(n,6n)$ & $6^{+3.3}_{-2.2}$ & $0.3^{+0.2}_{-0.1}$ & $0.005\pm0.003$ & $0^{+0.001}_{-0}$ & $0^{+0.001}_{-0}$ \\
 & & & & & & \\
 $^{73m}$Ge & $^{74}$Ge$(n,2n)$, $^{76}$Ge$(n,4n)$ & $1^{+1.9}_{-0.5}$ & $0.05^{+0.09}_{-0.020}$ & $0.38\pm0.21$ & $0.71\pm0.17$ & $0.70\pm0.17$\\
 & & & & & & \\
 $^{75m}$Ge & $^{76}$Ge$(n,2n)$ & $0^{+38}_{-0}$ & $0^{+1.7}_{-0.0}$ & $0.56\pm0.20$ & $0.96\pm0.2$ & $0.31\pm0.08$\\
 & & & & & & \\
 $^{77}$Ge & $^{76}$Ge$(n,\gamma)$ & $0^{+1.3}_{-0.0}$ & $0^{+0.3}_{-0.0}$ & $0.39\pm0.21$ & $0.021\pm0.005$ & $0.036\pm0.012$\\
 & & & & & & \\
 $^{77m}$Ge & $^{76}$Ge$(n,\gamma)$ & $0^{+23}_{-0}$ & $0^{+5.8}_{-0.0}$ & $0.39\pm0.21$ & $0.021\pm0.005$ & $0.016\pm0.007$\\
 & & & & & & \\ 
 \hline
 \hline
	\end{tabular}
\caption{Comparison of the detection rate from experiment, based on found candidate events in \DEM~data, and the simulation detection rate for different packages. The uncertainty for simulated values is given by the statistical error (68\%C.L.) of the simulation plus a 20\% uncertainty for the incoming muon flux as discussed in Ref.~\cite{Abgrall2017}.}
\label{table:ExpSim}
\end{table*}

\subsubsection{Distribution in time and energy}
\begin{figure}[t]
\centering
\includegraphics[width=0.85\columnwidth,keepaspectratio=true]{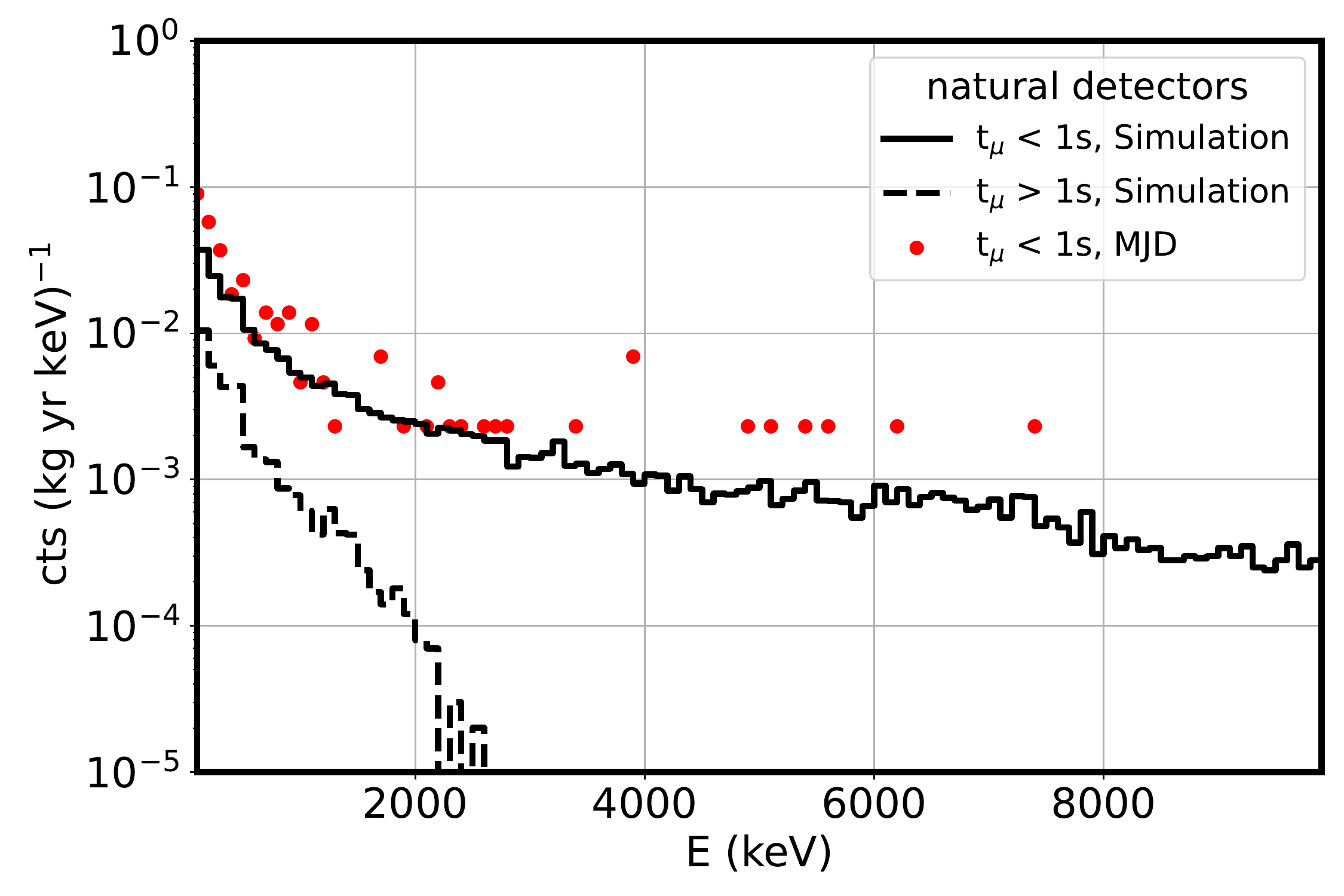}
\includegraphics[width=0.85\columnwidth,keepaspectratio=true]{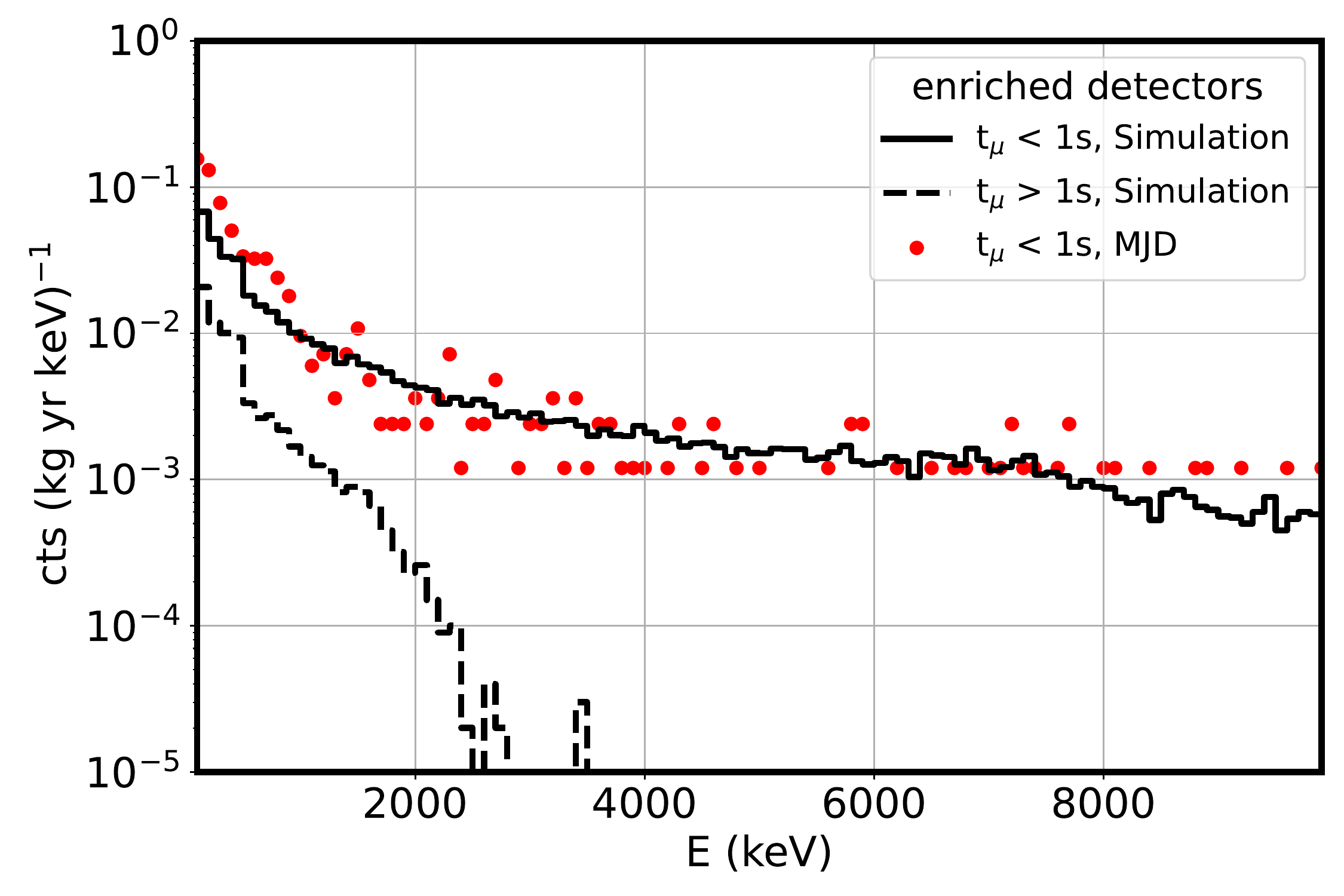}
\caption{(Color online) Comparison of the \DEM~data with simulations for natural (top) and enriched detectors (bottom) in 100\,keV binning. The red points represent \DEM~data in a one-second coincidence with the muon veto. The simulation by \MaGe~for the contribution of muon-induced events in the same time window is shown as well (black solid line). The simulated energy distribution for events that occur after one second in a single detector (black dashed) is mostly due to activation. No pulse shape cuts are applied for these distributions.} 
\label{fig:MJDcomparison}
\end{figure}

\begin{figure}[t]
\centering
\includegraphics[width=0.9\columnwidth,keepaspectratio=true]{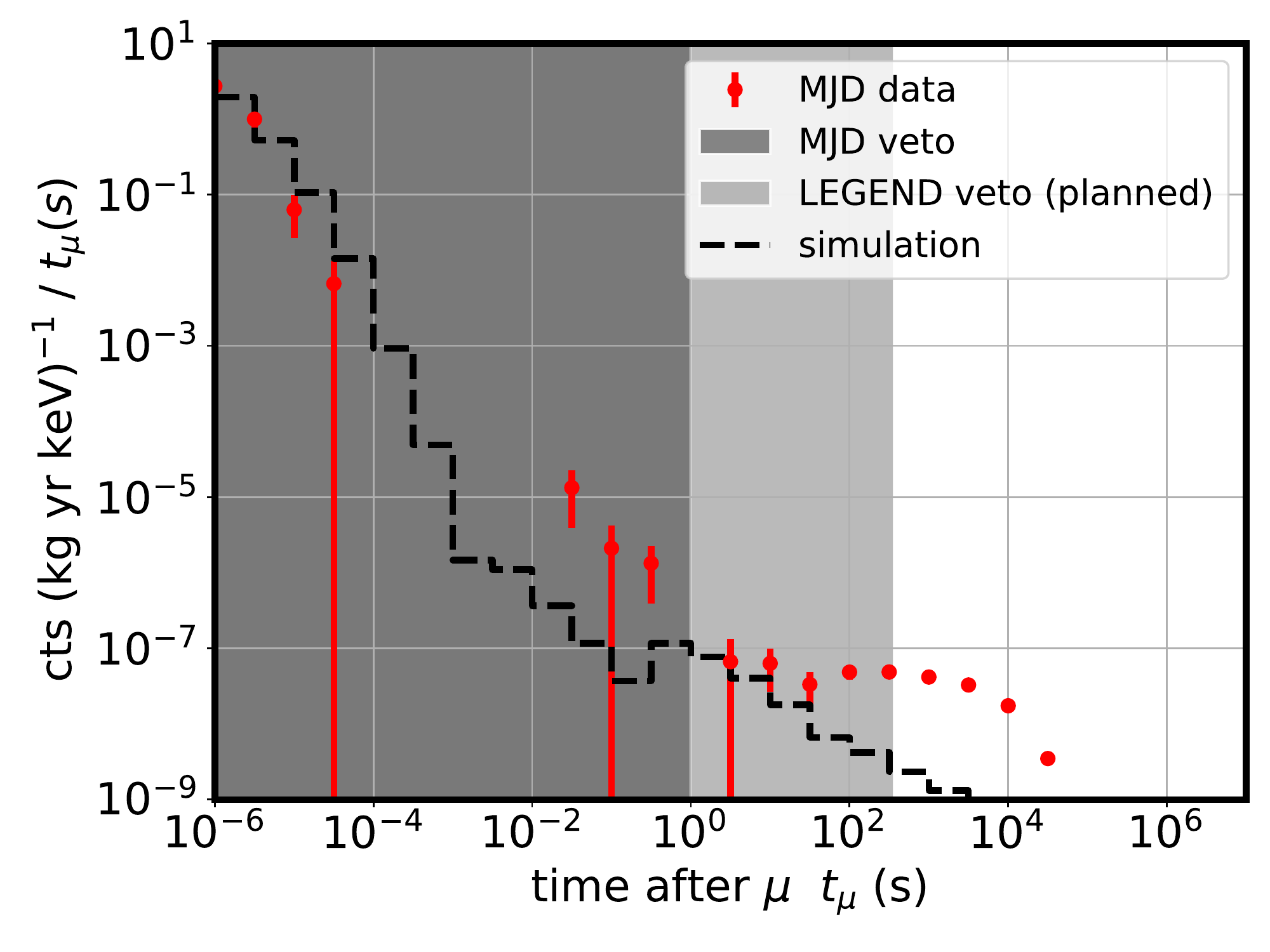}
\caption{(Color online) Time distribution of the events in the simulation between 1.5 and 2.5\,MeV for the enriched detectors (black dashed). The red dots represent data in the same window from \MJD~without any analysis cuts as shown in Ref. \cite{Alvis_2019}. The dark gray area shows events that occur within one second after an incident muon, which are removed by the current muon veto in the \DEM. The light gray area indicates the veto cut suggested in Ref.~\cite{Wiesinger_2018} for a future large-scale germanium experiment.}
\label{fig:IsotopesTime}
\end{figure}

As shown in Fig.~\ref{fig:MJDcomparison}, the energy distribution of events that are in coincidence with the muon veto is consistent in data and simulation. For \nonubb~analysis, the number of background events in the ROI is reduced when applying the veto. The remaining events contribute about 3$\times$10$^{-4}$ \cpKkgy~to the background around the Q-value in the enriched detectors. Table~\ref{table:data2} summarizes the simulated event rates of the isotopes which can decay and contribute to the ROI. For this summary, we considered events with energy deposits in the 400-keV wide window around the Q-value at 2.039\,MeV~\cite{Aalseth2018} that occur one second or later after the incident muon. Figure~\ref{fig:IsotopesTime} shows that the majority of muon-induced events which contribute to the \nonubb~ROI occur within this time. However, $\beta$-decaying isotopes, especially in decay chains involving multiple isotopes, can contribute at later times. Some events will contribute as background even after extended muon cuts like the one suggested by Ref.~\cite{Wiesinger_2018}. A comparison of experimental data in the ROI without any further analysis cuts indicates that simulation and experiment agree well for short time frames, as seen in Fig.~\ref{fig:IsotopesTime}. For longer times, when the correlation with the incident muon is not available, cosmogenic backgrounds in the ROI are subdominant. However, future experiments plan to lower background from construction material. This effectively reduces the dominant background sources while increasing the importance of the cosmogenic background. At the same time the experiment will be larger in size which allows the individual muons to interact with more germanium targets, so the importance of cosmogenic backgrounds will increase.

\begin{table*}
 \begin{tabular}{c|c|c|c|c}
 \hline 
 & \multicolumn{2}{c|}{\Geant 4.9.6} & \multicolumn{2}{c}{\Geant 4.10.5}\\
 Isotope & natural detectors & enriched detectors & natural detectors & enriched detectors \\
 & (10$^{-5}$\cpKkgy) & (10$^{-5}$\cpKkgy) & (10$^{-5}$\cpKkgy) & (10$^{-5}$\cpKkgy)\\
 \hline \hline
 & & & & \\
$^{58}$Co & 0.02 & $<$ 0.01 & $<$0.001 & 0.003 \\
$^{60}$Co & 0.09 & 0.01 & 0.09 & 0.04\\
& & & & \\
$^{61}$Cu & 0.02 & $<$ 0.01 & 0.02 & 0.02 \\
$^{62}$Cu & 0.17 & 0.08 & 0.03 & 0.03 \\
$^{66}$Cu & 0.22 & 0.16 & 0.01 & $<$0.013 \\
& & & & \\
$^{63}$Zn & 0.19 & $<$ 0.01 & $<$0.001 & $<$0.001 \\
$^{71}$Zn & 0.20 & 0.02 & $<$0.001 & $<$0.001 \\
$^{73}$Zn & 0.04 & 0.15 & $<$0.001 & 0.003 \\
& && & \\
$^{66}$Ga & 0.75 & 0.20 & $<$0.001 & $<$0.001\\ 
$^{68}$Ga & 4.94 & 0.27 & 0.28 & 0.25 \\
$^{72}$Ga & 0.28 & 1.07 & 0.58 & 0.65 \\
$^{74}$Ga & 0.03 & 0.11 & 0.23 & 0.36 \\
$^{75}$Ga & 2.19 & 1.18 & 0.42 & 0.43 \\
$^{76}$Ga & 0.05 & 0.19 & 0.01 & 0.02\\
& & & & \\
$^{66}$Ge & 0.03 & 0.01 & $<$0.001 & $<$0.001 \\
$^{67}$Ge & 0.60 & 0.15 & $<$0.001 & 0.07\\
$^{69}$Ge & 3.29 & 0.03 & $<$0.001 & $<$0.001\\
$^{77/77m}$Ge & 255 & 956 & 29.1 & 30.3\\
\hline
sum & 268 & 959 & 31 &	32 \\
\hline \hline
\end{tabular}
\caption{Event rates produced by the cosmogenic isotopes for events within the 400\,keV wide window around the Q-value~\cite{Aalseth2018} and occurring more than one second after the incident muon. No additional cuts on pulse shape are applied, see Fig.~\ref{fig:MJDcomparison}. One can assume a 100\% systematic uncertainty in the simulations, as discussed.}
\label{table:data2}
\end{table*}

\subsubsection{Uncertainty Discussion}
\label{sec:uncertainty}
\begin{figure}[t]
\centering
\includegraphics[width=0.95\columnwidth,keepaspectratio=true]{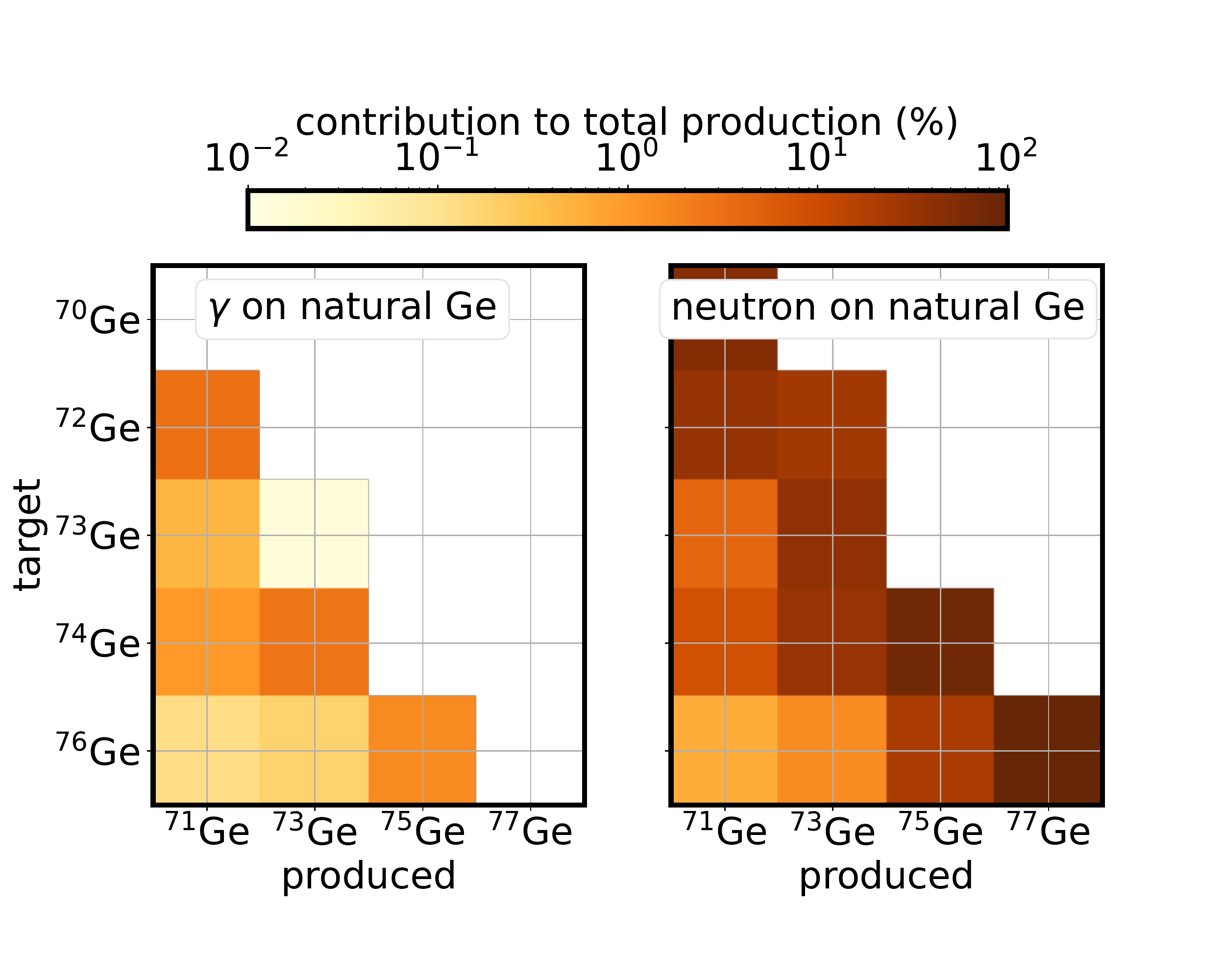}
\includegraphics[width=0.95\columnwidth,keepaspectratio=true]{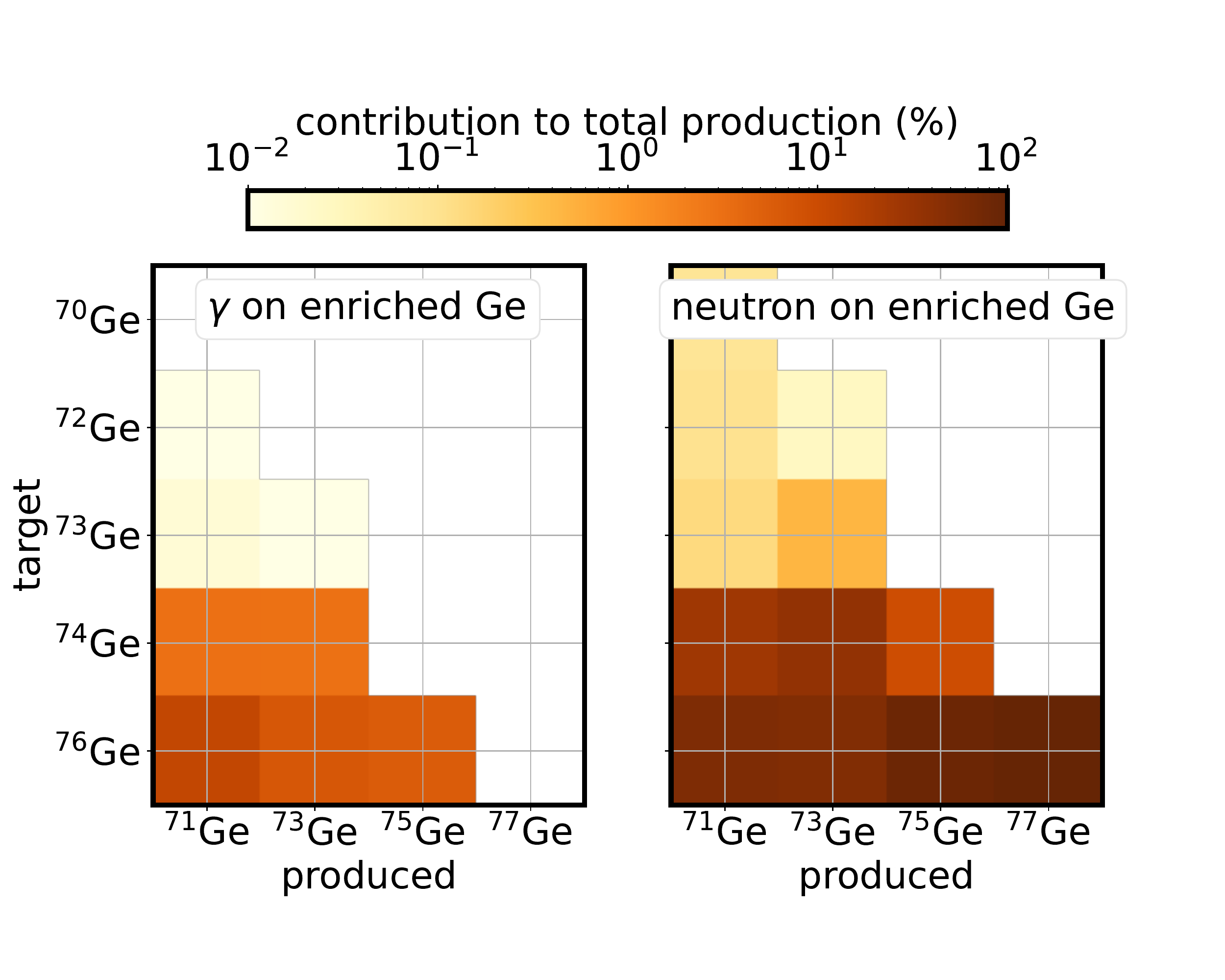}
\caption{(Color online) Contribution of each natural occurring isotope to the creation of the metastable states. The study is performed for naturally (top) and enriched (bottom) isotope mixtures, as given in Table~\ref{table:detisotopes}. The two channels $^{77}$Ge and $^{77m}$Ge are combined for this estimate since both are produced by capture on $^{76}$Ge.} 
\label{fig:Reaction}
\end{figure}

Other sources of background from natural radioactivity are neutrons produced by fission and ($\alpha$,n) processes in the rock. Reference~\cite{Akerib2015} estimated the integrated number of neutrons from these sources to be about a factor of 30 higher than those accompanying muons at the Davis Cavern at SURF. These neutrons have, as shown in Fig.~\ref{fig:NeutronShielding}, an energy distribution that reaches up into the MeV-range. Hence, their energies are too small to contribute to spallation processes which create the majority of the isotopes in Table~\ref{table:data2}. However, neutron capture reactions are possible. As discussed in the introduction, low-background experiments like the \DEM~consist of multiple shielding layers. Measurements and simulations~\cite{Barbagallo2020,Hu2017} indicate that the wall neutron flux is reduced by at least three orders of magnitude due to the combined 12-inch thick polyethylene layer and the 18-inch thick lead shield. Therefore, we expect a dominant production of slow neutrons by muons. This assumption is supported by the fact that we found no indication of prominent capture $\gamma$ rays from the copper which surrounds the detector. As stated, simulations have to cover a wide range of reaction cross-sections for various energies and isotopes. The simulations can be split into three major sections: 1) cosmogenic muons, with energies from a few GeV up to the TeV range and the creation of showers, 2) transport and interactions of a variety of particles in the accompanying shower, and 3) the decay of newly created radioactive isotopes. Several inputs can contribute to the total uncertainties of such a complex simulation framework. The uncertainty on the incoming muon rate is about 20\%~\cite{Abgrall2017} while the uncertainties on exposure are only about 2\%~\cite{Alvis_2019}. For this work, no further data cleaning cuts are applied in order to reduce the number of additional uncertainties. As shown in Fig.~\ref{fig:SimCodes}, the same geometry and input muon distributions will result in different rates in different reaction codes. Here, a large uncertainty comes from the physics models hidden in the simulation packages. Neutron physics often plays a special role since charged particles or photons can be shielded effectively with lead or other high-Z materials. As Table~\ref{table:ExpSim} shows, a large change has been observed between \Geant~versions. One contributing factor is the use of the evaluated data tables in the newer version, which aims to improve the predictive power of the simulation package \cite{ALLISON2016186}. The predicted number of events in the newer version of \Geant~is also consistent with the FLUKA physics, which supports these changes. Various simulation packages use slightly different neutron physics models. Databases for neutron cross-sections are often incomplete, or only exist for energies and materials relevant to reactors. This problem was noted previously and comparisons between packages have been done to study neutron propagation or muon-induced neutron production~\cite{Araujo2005,Yeh2007}. 
The influence of the isotope mixture and its uncertainty on the final results was investigated as well. Given the intense CPU-time needed for the as-built \DEM\, simulation, a simplified calculation was done to estimate the dominant reaction channels. From \MaGe, the flux of neutrons and $\gamma$ rays inside the innermost cavity was tabulated and folded with the isotopic abundance as given in Table~\ref{table:detisotopes} as well as the reaction cross section calculated by TALYS \cite{Koning2012,TALYS2018}. As shown in Fig.~\ref{fig:Reaction}, neutrons are the dominating projectiles to create the meta-stable isomers used in this study. For a natural isotope composition neutron capture reactions dominate the production over knockout reactions like ($\gamma,n$) or ($n,2n$). Since the natural isotope composition is well understood only minor uncertainties are introduced. For enriched detectors, knockout reactions as listed in Table~\ref{table:ExpSim} dominate the production mechanisms. Hence, the lighter germanium isotopes and their large relative uncertainties only contribute on a negligible scale. 

In the current-generation experiments, the cosmogenic backgrounds are only a small background contribution since the total background is on the order of 4.7$\times$10$^{-3}$\,\cpKkgy~for \MJD~\cite{Alvis_2019}, and 5.6$\times$10$^{-4}$\,\cpKkgy~for \Gerda~\cite{PhysRevLett.120.132503,Agostini1445}. Due to the different shielding approach, the \Gerda\ background contribution by cosmogenics can not be compared directly to the \MJD. This will be discussed in the next section. However, in order to improve the background rate for next generation experiments, a detailed understanding of the cosmogenic backgrounds becomes necessary~\cite{LEGEND2021}.

\section{Outlook to a Ge-based tonne-scale \nonubb~effort}
\begin{figure}[t]
\centering
\includegraphics[width=0.9\columnwidth,keepaspectratio=true]{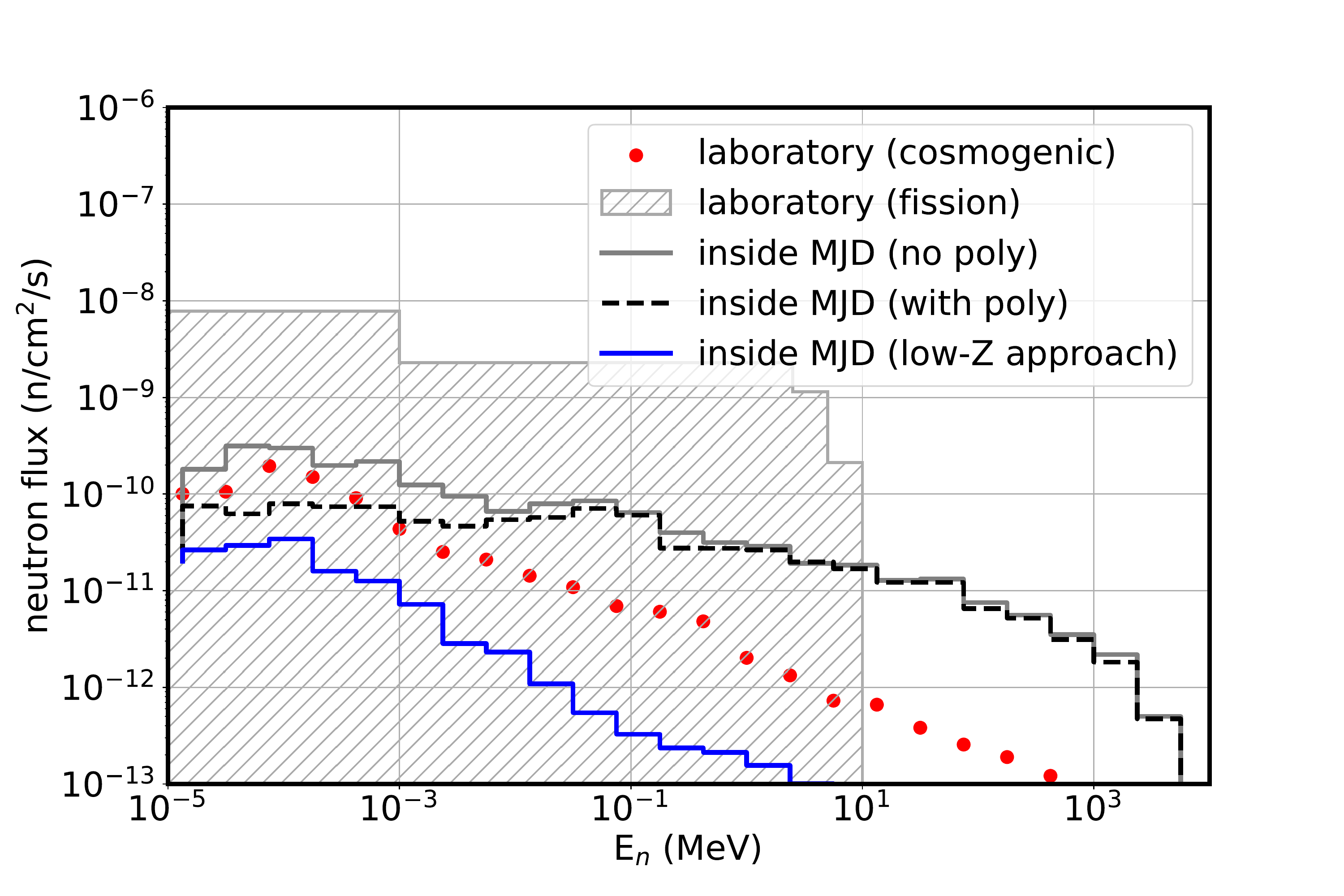}
\caption{(Color online) Neutron flux at the 4850\,ft level for various shielding scenarios. The red dots and the grey area curve show the neutron flux entering the experimental cavity from cosmogenics and due to fission in the rock \cite{Akerib2015}. The increase in flux after the innermost shielding layer of the \DEM~ (black dashed) is due to the production of additional neutrons by muons in lead. Different shielding approaches, e.g. no poly-shield (grey), or low-Z approach with liquid argon (blue) can affect the flux.}
\label{fig:NeutronShielding}
\end{figure}

The results in Fig.~\ref{fig:MJDcomparison} suggest that simulations are capable of qualitatively describing the cosmogenic contribution to the background budget. However, as shown in Fig.~\ref{fig:SimCodes}, uncertainties can become a problem and even more prominent when discussing the background of a tonne-scale \nonubb~experiment, such as the LEGEND experiment~\cite{LEGEND2021}. The sensitivities for next-generation efforts are strongly dependent on the background level~\cite{Agostini2017, LEGEND2021}. If the background is “zero”, the sensitivity scales linearly with the exposure; otherwise, the sensitivity only scales as the square root of the exposure. For LEGEND-1000, the goal is to reduce the background to 10$^{-5}$\,\cpKkgy. Hence, the integrated rates in Table~\ref{table:data2} would be too high for the background in the future experiment. As shown in Fig.~\ref{fig:IsotopesTime}, one can increase the veto time after each muon in order reduce the background, but this technique is limited and increases the amount of detector dead time, especially for underground laboratories with less rock overburden and consequently higher muon flux. The design and the location of the tonne-scale experiment directly impact the background budget with respect to cosmogenic contributions. One major feature of the next-generation design is the usage of low-Z shielding material, such as the liquid argon shield in \Gerda. In addition to its active veto capability, argon as a shielding material directly affects the secondary neutron production close by the germanium crystals. Figure~\ref{fig:NeutronShielding} shows that the neutron flux at the 4850\,ft level in simulations can change as the shielding configuration changes. The total neutron flux entering the cavity from the current simulation is estimated to be $(0.78\pm 0.16) \times 10^{-9}$\,n\,cm$^{-2}$\,s$^{-1}$ which is in reasonable agreement with previous predictions by Mei-Hime \cite{MeiHime2006} $(0.46\pm 0.10) \times 10^{-9}$\,n\,cm$^{-2}$\,s$^{-1}$, and an estimate by the LUX collaboration ~\cite{Akerib2015} $(0.54\pm 0.01)\times 10^{-9}$\,n\,cm$^{-2}$\,s$^{-1}$. The installation of the 30-cm thick poly-shield suppresses the low-energy portion of the neutron flux while the high-energy portion of the neutron flux is mostly unaffected. This is because most of the fast secondary neutron flux is produced inside the lead shielding. To understand the effect of a low-Z shielding material, the 18-inch thick lead shield in the \DEM~simulations was replaced with a 4.4-meter thick liquid argon shield. This thickness results in the same suppression factor for 2.6\,MeV $\gamma$ rays. In the simulations, this liquid argon shield suppresses the neutron flux inside the inner-most shielding. An instrumented liquid argon shield can further suppress delayed signatures, reducing the total cosmogenic contribution. As shown in Table~\ref{table:data2}, $^{77}$Ge, the main contribution to the ROI, is mostly created by low-energy neutron capture which would be suppressed by a liquid argon shield. Table~\ref{table:observed_events_ROI} shows the background estimation for a \DEM-scale experiment with different shield configurations. The 1-sec muon veto can suppress the muon-induced background by roughly a factor of ten; however, the liquid argon shield can further reduce the background. 
In a tonne-scale experiment with \DEM-style shielding at 4850-ft depth, the current cosmogenic background rate shown in Table~\ref{table:data2} represents 200\% of the background budget for LEGEND-1000. However, a low-Z shielding approach, as well as analysis cuts as given in Ref.~\cite{Wiesinger_2018} drop this number to the percent level. Especially time and spatial correlations, see Ref.~\cite{Ejiri2005}, are very effective in reducing the effects of correlated signals from cosmogenic particles deep underground. As shown in Ref.\,\cite{LEGEND2021} a deeper laboratory will reduce the cosmogenic background, as it scales with the muon flux at the first order. However, details like shielding materials, additional neutron absorbers, detector arrangement, and analysis cuts help to reduce the contribution.

\begin{table*}
\begin{center}
 \begin{tabular}{c|c|c}
 	\hline
 	 &\multicolumn{2}{c}{Rate}\\
 	 &\multicolumn{2}{c}{10$^{-5}$\cpKkgy}\\\hline
	 &Natural & Enriched \\
	\hline
	\hline 
	\textbf{lead shield (no poly)}& &\\
	 total & 712 & 460 \\
	 $1$ s muon veto & 53 & 59 \\\hline
 \textbf{lead shield (with poly)}& &\\
 total & 424 & 260 \\
	 $1$ s muon veto & 27 & 32 \\\hline
 \textbf{liquid Argon}& &\\
 total & 12.6 & 7.9 \\
	 $1$ s muon veto & 0.9 & 1.8 \\
	 delayed tag \cite{Wiesinger_2018} & 0.09 & 0.18\\
	\hline\hline	
 \end{tabular}
 \end{center}
 \caption{Cosmogenic event rate in the 400-keV wide window at the Q-Value for lead and liquid argon shielding options at the 4850\,ft level of SURF, without additional pulse shape analysis. For lead shielding, the two cases in Fig.~\ref{fig:NeutronShielding} are shown representing the two extremes during the \DEM~construction: without the poly shield at the beginning and with the 30-cm thick poly in the final configuration. }
 \label{table:observed_events_ROI}
\end{table*}

\section{Summary}
This work presents a search for cosmogenically produced isotopes in the \MJD~and compares the detected number to predictions from simulations. The number of isotopes agrees reasonably well, and the overall distribution in energy and time are in good agreement to measured distributions. However, differences between simulation packages lead to uncertainties that are not negligible. Given the complexity of the simulations, uncertainties of a factor of two or more should be considered. It has been shown that for a future Ge-based tonne-scale experiment, the design directly affects the production of isotopes and the background to the ROI. Low-Z shielding like liquid argon in combination with analysis cuts can have similar impact as a deeper laboratory when reducing the effect of cosmogenic radiation.

\section{Acknowledgements}
This material is based upon work supported by the U.S.~Department of Energy, Office of Science, Office of Nuclear Physics under contract / award numbers DE-AC02-05CH11231, DE-AC05-00OR22725, DE-AC05-76RL0130, DE-FG02-97ER41020, DE-FG02-97ER41033, DE-FG02-97ER41041, DE-SC0012612, DE-SC0014445, DE-SC0018060, and LANLEM77/LANLEM78. We acknowledge support from the Particle Astrophysics Program and Nuclear Physics Program of the National Science Foundation through grant numbers MRI-0923142, PHY-1003399, PHY-1102292, PHY-1206314, PHY-1614611, PHY-1812409, and PHY-1812356. We gratefully acknowledge the support of the Laboratory Directed Research \& Development (LDRD) program at Lawrence Berkeley National Laboratory for this work. We gratefully acknowledge the support of the U.S.~Department of Energy through the Los Alamos National Laboratory LDRD Program and through the Pacific Northwest National Laboratory LDRD Program for this work. We gratefully acknowledge the support of the South Dakota Board of Regents Competitive Research Grant. We acknowledge support from the Russian Foundation for Basic Research, grant No.~15-02-02919. We acknowledge the support of the Natural Sciences and Engineering Research Council of Canada, funding reference number SAPIN-2017-00023, and from the Canada Foundation for Innovation John R.~Evans Leaders Fund. This research used resources provided by the Oak Ridge Leadership Computing Facility at Oak Ridge National Laboratory and by the National Energy Research Scientific Computing Center, a U.S.~Department of Energy Office of Science User Facility. We thank our hosts and colleagues at the Sanford Underground Research Facility for their support.
\bibliography{neutron}
\end{document}